\documentclass[12pt,twoside]{article}
\usepackage{correl}

\def\TheTitle{Orbital Physics}
\def\TheHeading{Orbital Physics}
\def\TheAuthor{Andrzej M. Ole\'{s}}
\def\TheInstitute{Marian Smoluchowski Institute of Physics \\ Jagiellonian University}
\def\TheAddress{Prof. S. {\L}ojasiewicza 11, Krak\'ow, Poland}
\def\TheChapter{11}                       

\newcommand{\lla}{\left\langle}
\newcommand{\rra}{\right\rangle}
\newcommand{\bec}{\begin{center}}
\newcommand{\enc}{\end{center}}
\newcommand{\beq}{\begin{equation}}
\newcommand{\eeq}{\end{equation}}
\newcommand{\bea}{\begin{eqnarray}}
\newcommand{\eea}{\end{eqnarray}}
\newcommand{\nn}{\nonumber}

\begin{document}
\MakeTitle           

\section{Introduction: strong correlations at orbital degeneracy}

Strong local Coulomb interactions lead to electron localization in Mott
or charge transfer correlated insulators. The simplest model of a Mott
insulator is the non-degenerate Hubbard model where large intraorbital
Coulomb interaction $U$ suppresses charge fluctuations due to the
kinetic energy $\propto t$.
As a result, the physical properties of a Mott insulator are
determined by an interplay of kinetic exchange $\propto J$, with
\index{$t$-$J$ model}
\index{Mott insulator}
\beq
\label{J}
J=\frac{4t^2}{U},
\eeq
derived from the Hubbard model at $U\gg t$, and the motion of holes in
the restricted Hilbert space without double occupancies, as described
by the $t$-$J$ model \cite{Cha77}. Although this generic model captures
the essential idea of strong correlations, realistic correlated
insulators arise in transition metal oxides (or fluorides) and the
degeneracy of partly filled and nearly degenerate $3d$ (or $4d$)
strongly correlated states has to be treated explicitly.
Quite generally, strong local Coulomb interactions lead then to the
multitude of quite complex behavior with often puzzling transport
and magnetic properties \cite{Ima98}. The theoretical understanding
of this class of compounds, including the colossal magneto-resistance
(CMR) manganites as a prominent example \cite{Dag01}, has to include
not only spins and holes but in addition orbital degrees of freedom
which have to be treated on equal footing with the electron spins
\cite{Kug82}. For a Mott insulator with transition metal ions in $d^m$
configurations, charge excitations along the bond $\langle ij\rangle$,
$d_i^md_j^m\rightleftharpoons d_i^{m+1}d_j^{m-1}$,
lead to spin-orbital superexchange which couples two
neighboring ions at sites $i$ and~$j$.

\begin{figure}[t!]
\bec
\includegraphics[width=11cm]{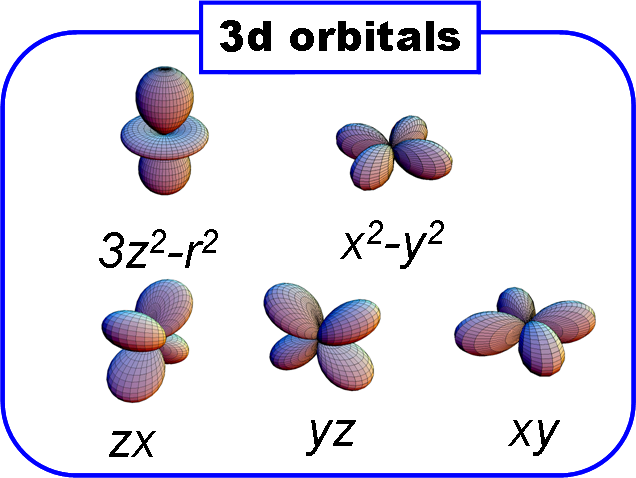}
\enc
\caption{
Schematic representation of $3d$ orbitals:
Top --- two $e_g$ orbitals $\{3z^2-r^2,x^2-y^2\}$;
Bottom --- three $t_{2g}$ orbitals $\{zx,yz,xy\}$.
Image by courtesy of Yoshinori Tokura.
}
\label{fig:3d}
\end{figure}

It is important to realize that modeling of transition metal oxides
can be performed on different levels of sophistication. We shall
present some effective orbital and spin-orbital superexchange models
for the correlated $3d$-orbitals depicted in Fig.~\ref{fig:3d}
coupled by hopping $t$ between
nearest neighbor ions on a perovskite lattice, while the hopping for
other lattices may be generated by the general rules formulated by
Slater and Koster \cite{Sla54}. The orbitals have particular shapes
and belong to two irreducible
representations of the $O_h$ cubic point group: \hfill \\
($i$)~a two-dimensional (2D) representation of $e_g$-orbitals
$\{3z^2-r^2,x^2-y^2\}$, and \hfill \\
($ii$)~a three-dimensional (3D) representation of $t_{2g}$-orbitals
$\{xy,yz,zx\}$. \hfill\\
In case of absence of any tetragonal distortion or crystal-field due to
surrounding oxygens, the $3d$-orbitals are degenerate within each
irreducible representation of the $O_h$ point group and have typically
a large splitting $10D_q\sim~2.0$ eV between them. When such degenerate
orbitals are party filled, electrons (or holes) have both spin and
orbital degree of freedom. The kinetic energy $H_t$ in a perovskite
follows from the hybridization between $3d$ and $2p$-orbitals. In an
effective $d$-orbital model the oxygen $2p$-orbitals are not included
explicitly and we define the hopping element $t$ as the largest hopping
element obtained for two orbitals of the same type which belong to the
nearest neighbor $3d$ ions.

We begin with conceptually simpler $t_{2g}$ orbitals where finite
hopping $t$ results from the ${d-p}$ hybridization along $\pi$-bonds and
couples each a pair of identical orbitals active along a given bond.
Each $t_{2g}$ orbital is active along two cubic axes and the hopping is
forbidden along the one perpendicular to the plane of this orbital,
e.g. the hopping between two $xy$-orbitals is not allowed along the $c$
axis (due to the cancelations caused by orbital phases).
It is therefore convenient to introduce the following short-hand
notation for the orbital degree of freedom \cite{Kha00},
\begin{equation}
|a\rangle \equiv |yz\rangle, \hskip .7cm
|b\rangle \equiv |zx\rangle, \hskip .7cm
|c\rangle \equiv |xy\rangle.
\label{t2g}
\end{equation}
The labels $\gamma=a,b,c$ thus refer to the cubic axes where the
hopping is absent for orbitals of a given type,
\begin{equation}
H_t(t_{2g})=-t \sum_{\alpha}\sum_{\langle ij\rangle\parallel\gamma\neq\alpha}
    a_{i\alpha\sigma}^{\dagger}a_{j\alpha\sigma}^{},
\label{Htg}
\end{equation}
Here $a^{\dagger}_{i\alpha\sigma}$ is an electron creation operator in
a $t_{2g}$-orbital $\alpha\in\{xy,yz,zx\}$ with spin
$\sigma=\uparrow,\downarrow$ at site $i$, and the local electron
density operator for a spin-orbital state is
$n_{i\alpha\sigma}^{}=a^{\dagger}_{i\alpha\sigma}a^{}_{i\alpha\sigma}$.
Not only spin but also orbital flavor is conserved in the hopping
process $\propto t$.

The hopping Hamiltonian for $e_g$ electrons couples two directional
$e_g$-orbitals $\{|i\zeta_{\alpha}\rangle,|i\zeta_{\alpha}\rangle\}$
along a $\sigma$-bond $\lla ij\rra$ \cite{Fei05},
\begin{equation}
H_t(e_g)=-t \sum_{\alpha}\sum_{\langle ij\rangle\parallel\alpha,\sigma}
    a_{i\zeta_{\alpha}\sigma}^{\dagger}a_{j\zeta_{\alpha}\sigma}^{}.
\label{H_zeta}
\end{equation}
Indeed, the hopping with amplitude $-t$ between sites $i$ and $j$
occurs only when an electron with spin $\sigma$ transfers between two
directional orbitals $|\zeta_{\alpha}\rangle$ oriented along the bond
$\langle ij\rangle$ direction, i.e.,
$|\zeta_{\alpha}\rangle\propto 3x^2-r^2$, $3y^2-r^2$, and $3z^2-r^2$
along the cubic axis $\alpha=a$, $b$, and $c$. We will similarly denote
by $|\xi_{\alpha}\rangle$ the orbital which is orthogonal to
$|\zeta_{\alpha}\rangle$ and is oriented perpendicular to the bond
$\langle ij\rangle$ direction, i.e.,
$|\xi_{\alpha}\rangle\propto y^2-z^2$, $z^2-x^2$, and $x^2-y^2$ along
the axis $\alpha=a$, $b$, and $c$. For
a moment we consider only electrons with one spin, $\sigma=\uparrow$,
to focus on the orbital problem. While such a choice of an
over-complete basis $\{\zeta_a,\zeta_b,\zeta_c\}$ is convenient for
writing down the kinetic energy, a particular orthogonal basis is
needed. The usual choice is to take
\begin{equation}
\label{real}\textstyle{
|z\rangle\equiv \frac{1}{\sqrt{6}}(3z^2-r^2),
\hspace{0.7cm}
|\bar{z}\rangle\equiv \frac{1}{\sqrt{2}}(x^2-y^2),}
\end{equation}
called {\it real $e_g$ orbitals\/} \cite{Fei05}.
However, this basis is the natural one only for the bonds parallel to
the $c$ axis but not for those in the $(a,b)$ plane, and for
$\uparrow$-spin electrons the hopping reads (here for clarity we omit
spin index $\sigma$)
\beq
\label{Hxz}
H_t^{\uparrow}(e_g)= -\frac{1}{4}t\!\sum_{\langle ij\rangle\parallel ab}\!
  \left[3a_{i\bar{z}}^{\dagger}a_{j\bar{z}}^{}
  +a_{iz}^{\dagger}a_{jz}^{}
\mp\sqrt{3}\left(a_{i\bar{z}}^{\dagger}a_{jz}^{}
+a_{iz}^{\dagger}a_{j\bar{z}}^{}\right)\right]
 -t\sum_{\langle ij\rangle\parallel c} a_{iz}^{\dagger}a_{jz}^{},
\eeq
and although this expression is of course cubic invariant, it does not
manifest this symmetry but takes a very different appearance depending
on the bond direction. However, the symmetry is better visible using
the basis of {\it complex $e_g$ orbitals} at each site $j$ \cite{Fei05},
\begin{equation}
\label{complex}
\textstyle{
|j+\rangle=\frac{1}{\sqrt{2}}\big(|jz\rangle - i|j\bar{z}\rangle\big),
\hspace{0.7cm}
|j-\rangle=\frac{1}{\sqrt{2}}\big(|jz\rangle + i|j\bar{z}\rangle\big),}
\end{equation}
corresponding to ``up'' and``down'' pseudospin flavors, with the local
pseudospin operators defined as
\beq
\tau_i^+ \equiv c_{i +}^{\dagger} c_{i -}^{},    \hspace{.7cm}
\tau_i^- \equiv c_{i -}^{\dagger} c_{i +}^{},    \hspace{.7cm}
\tau_i^z \equiv\textstyle{\frac{1}{2}}
  (c_{i +}^{\dagger} c_{i +}^{} - c_{i -}^{\dagger} c_{i -}^{} )
= \textstyle{\frac{1}{2}}(n_{i +} - n_{i -}).
\label{pseudospin}
\eeq
The three directional $|i\zeta_{\alpha}\rangle$ and three planar
$|i\xi_{\alpha}\rangle$ orbitals at site $i$, associated with the three
cubic axes ($\alpha=a$, $b$, $c$), are the real orbitals,
\begin{eqnarray}
\label{zeta}
|i \zeta_{\alpha}\rangle &=&\textstyle{\frac{1}{\sqrt{2}}}\,
    \left[e^{-i\vartheta_{\alpha}/2}|i+\rangle
    +e^{+i\vartheta_{\alpha}/2}|i-\rangle\right]
  = \cos(\vartheta_{\alpha}/2)|iz\rangle
   -\sin(\vartheta_{\alpha}/2)|i\bar{z}\rangle,   \\
\label{xi}
|i   \xi_{\alpha}\rangle &=&\textstyle{\frac{1}{\sqrt{2}}}\,
    \left[e^{-i\vartheta_{\alpha}/2}|i+\rangle
    -e^{+i\vartheta_{\alpha}/2}|i-\rangle\right]
  = \sin(\vartheta_{\alpha}/2)|iz\rangle
   +\cos(\vartheta_{\alpha}/2)|i\bar{z}\rangle,
\end{eqnarray}
with the phase factors
$\vartheta_{ia}=-4\pi/3$, $\vartheta_{ib}=+4\pi/3$, and
$\vartheta_{ic}=0$, and thus correspond to the pseudospin lying in the
equatorial plane and pointing in one of the three equilateral ``cubic''
directions defined by the angles $\{\vartheta_{i\alpha}\}$.

Using the above complex-orbital representation (\ref{complex}) we can
write the {\it orbital Hubbard model\/} for $e_g$ electrons with only
one spin flavor $\sigma=\uparrow$ in a form similar to the spin Hubbard
model,
\index{$e_g$ orbital Hubbard model}
\beq
\label{egHub}
{\cal H}^{\uparrow}(e_g)= -\frac{1}{2} t \sum_{\alpha}\!
\sum_{\langle ij\rangle\parallel\alpha}
  \Big[\Big(a_{i+}^{\dagger}a_{j+}^{}+a_{i-}^{\dagger}a_{j-}^{}\Big)
  + \gamma \Big(e^{-i\chi_{\alpha}}a_{i+}^{\dagger}a_{j-}^{}
    +e^{+i\chi_{\alpha}}a_{i-}^{\dagger}a_{j+}^{}\Big)\Big]
  +\,\bar{U} \sum_in_{i+}^{}n_{i-}^{},
\eeq
with $\chi_a=+2\pi/3$, $\chi_b=-2\pi/3$, and $\chi_c=0$, and where the
parameter $\gamma$, explained below, takes for $e_g$ orbitals the value
$\gamma=1$. The appearance of the phase factors $e^{\pm i\chi_{\alpha}}$
is characteristic of the orbital problem --- they occur because the
orbitals have an actual shape in real space so that each hopping process
depends on the bond direction and may change the orbital flavor.
The interorbital Coulomb interaction $\propto\bar{U}$
couples the electron densities in basis orbitals
${n_{i\alpha}=a_{i\mu}^{\dagger}a_{i\mu}^{}}$, with $\mu\in\{+,-\}$; its
form in invariant under any local basis transformation to a pair of
orthogonal orbitals, i.e., it gives an energy $\bar{U}$ for a double
occupancy either when two real orbitals are simultaneously occupied,
$\bar{U}\sum_in_{iz}n_{i\bar{z}}$, or when two complex orbitals are
occupied, $\bar{U}\sum_in_{i+}n_{i-}$.

In general, on-site Coulomb interactions between two interacting
electrons in $3d$-orbitals depend both on spin and orbital indices and
the interaction Hamiltonian takes the form of the degenerate Hubbard
model. Note that the electron interaction parameters in this model are
effective ones, i.e., the $2p$-orbital parameters of O (F) ions
renormalize on-site Coulomb interactions for $3d$-orbitals. The general
form which includes only two-orbital interactions and the anisotropy of
Coulomb and exchange elements is \cite{Ole05}:
\begin{eqnarray}
\label{Hub}
H_{int}&=&
   U\sum_{i\alpha}n_{i\alpha  \uparrow}n_{i\alpha\downarrow}
+\sum_{i,\alpha<\beta}\left(U_{\alpha\beta}-\frac{1}{2}J_{\alpha\beta}\right)\,
                    n_{i\alpha}n_{i\beta}
-2\sum_{i,\alpha<\beta}J_{\alpha\beta}\,
    {\vec S}_{i\alpha}\cdot{\vec S}_{i\beta}      \nonumber \\
&+& \sum_{i,\alpha<\beta}J_{\alpha\beta}
\left(a^{\dagger}_{i\alpha\uparrow}  a^{\dagger}_{i\alpha\downarrow}
      a^{       }_{i\beta\downarrow} a^{       }_{i\beta\uparrow}
     +a^{\dagger}_{i\beta\uparrow}   a^{\dagger}_{i\beta\downarrow}
      a^{       }_{i\alpha\downarrow}a^{       }_{i\alpha\uparrow}\right).
\end{eqnarray}
\index{on-site Coulomb interactions}
Here $a^{\dagger}_{i\alpha\sigma}$ is an electron creation operator in
any $3d$-orbital $\alpha\in\{xy,yz,zx,3z^2-r^2,x^2-y^2\}$ and
$\bar{\sigma}equiv -\sigma$,
with spin states $\sigma=\uparrow,\downarrow$ at site $i$.
The parameters $\{U,U_{\alpha\beta},J_{\alpha\beta}\}$ depend in the
general case on the three Racah parameters $A$, $B$ and $C$ \cite{Griff}
which may be derived from somewhat screened atomic values. While the
intraorbital Coulomb element is identical for all $3d$-orbitals,
\begin{equation}
\label{U}
  U=A+4B+3C,                                                   \\
\end{equation}
the interorbital Coulomb $U_{\alpha\beta}$ and exchange
$J_{\alpha\beta}$ elements are anisotropic and depend on the involved
pair of orbitals; the values of $J_{\alpha\beta}$ are given in
Table~\ref{tab:uij}. The Coulomb $U_{\alpha\beta}$ and exchange
$J_{\alpha\beta}$ elements are related to the intraorbital element $U$
by a relation which guarantees the invariance of interactions in the
orbital space,
\begin{equation}
\label{Uab}
  U=U_{\alpha\beta}+2J_{\alpha\beta}.
\end{equation}

\begin{table}[b!]
\caption{
On-site interorbital exchange elements $J_{\alpha\beta}$ for $3d$
orbitals as functions of the Racah parameters $B$ and $C$ (for more
details see Ref. \cite{Griff}). }
\vskip .2cm
\bec
\begin{tabular}{cccccc}
\hline\hline
$3d$ orbital& $xy$   &  $yz$  & $zx$   & $x^2-y^2$ & $3z^2-r^2$ \cr
\hline
  $xy$      & $0$    & $3B+C$ & $3B+C$ &    $C$    & $4B+C$  \cr
  $yz$      & $3B+C$ & $0$    & $3B+C$ & $3B+C$    &  $B+C$  \cr
  $zx$      & $3B+C$ & $3B+C$ & $0$    & $3B+C$    &  $B+C$  \cr
 $x^2-y^2$  &    $C$ & $3B+C$ & $3B+C$ &  $0$      & $4B+C$  \cr
 $3z^2-r^2$ & $4B+C$ &  $B+C$ &  $B+C$ & $4B+C$    &  $0$    \cr
\hline\hline
\end{tabular}
\enc
\label{tab:uij}
\end{table}

In all situations where only the orbitals belonging to a single irreducible
representation of the cubic group ($e_g$ or $t_{2g}$) are partly filled,
as e.g. in the titanates, vanadates, nickelates, or copper fluorides,
the filled (empty) orbitals do not contribute, and the relevant exchange
elements $J_{\alpha\beta}$ are all the same (see Table~\ref{tab:uij}),
i.e., for $t_{2g}$ ($e_g$) orbitals,
\index{Hund's exchange}
\bea
\label{JHt}
J_H^t&=&3B+C, \\
\label{JHe}
J_H^e&=&4B+C.
\eea
Then one may use a simplified \textit{degenerate} Hubbard model with
isotropic form of on-site interactions
(for a given subset of $3d$ orbitals) \cite{Ole83},
\index{degenerate Hubbard model}
\begin{eqnarray}
\label{Hee}
H_{int}^{(0)}&=&
   U\sum_{i\alpha}n_{i\alpha  \uparrow}n_{i\alpha\downarrow}
 +\left(U-\frac{5}{2}J_H\right)\sum_{i,\alpha<\beta}n_{i\alpha}n_{i\beta}
 - 2J_H\sum_{i,\alpha<\beta}\,\vec{S}_{i\alpha}\cdot\vec{S}_{i\beta}
                                                          \nonumber \\
&+& J_H\sum_{i,\alpha<\beta}
\left(a^{\dagger}_{i\alpha\uparrow}  a^{\dagger}_{i\alpha\downarrow}
      a^{       }_{i\beta\downarrow} a^{       }_{i\beta\uparrow}
     +a^{\dagger}_{i\beta\uparrow}   a^{\dagger}_{i\beta\downarrow}
      a^{       }_{i\alpha\downarrow}a^{       }_{i\alpha\uparrow}\right).
\end{eqnarray}
It has two Kanamori parameters:
the Coulomb intraorbital element $U$ (\ref{U}) and Hund's exchange $J_H$
standing either for $J_H^t$ (\ref{JHt}) or for $J_H^e$ (\ref{JHe}).
Now $\bar{U}\equiv(U-3J_H)$ in Eq. (\ref{egHub}).
We emphasize that in a general case when both types of orbitals are
partly filled (as in the CMR manganites) and both thus participate in
charge excitations, the above Hamiltonian with a single Hund's exchange
element $J_H$ is insufficient and the full anisotropy given in Eq.
(\ref{Hee}) has to be used instead to generate correct charge
excitation spectra of a given transition metal ion \cite{Griff}.

\section{Orbital and compass models}
\label{sec:orbi}

If the spin state is ferromagnetic (FM) as e.g. in the $ab$ planes of
KCuF$_3$ (or LaMnO$_3$), charge excitations
$d_i^md_j^m\rightleftharpoons d_i^{m+1}d_j^{m-1}$ with $m=9$ (or $m=4$)
concern only high-spin (HS) $^3A_1$ (or~$^6A_1$) state and the
superexchange interactions reduce to an orbital superexchange
model~\cite{vdB99}. Thus we begin with an orbital model for $e_g$-holes
in K$_2$CuF$_4$, with a local basis at site $i$ defined by two real
$e_g$-orbitals, see Eq. (\ref{real}),
being a local $e_g$-orbital basis at each site. The basis consists of
a directional orbital $|i\zeta_c\rangle\equiv|iz\rangle$ and the
planar orbital $|i\xi_c\rangle\equiv|i\bar{z}\rangle$. Other
equivalent orbital bases are obtained by rotation of the above pair
of orbitals by angle $\vartheta$ to
\bea
|i\vartheta\rangle&=&
 \cos\left(\vartheta/2\right)|iz\rangle
-\sin\left(\vartheta/2\right)|i\bar{z}\rangle, \nonumber\\
|i\bar{\vartheta}\rangle&=&
 \sin\left(\vartheta/2\right)|iz\rangle
+\cos\left(\vartheta/2\right)|i\bar{z}\rangle,
\eea
i.e., to a pair $\{|i\vartheta\rangle,|i,\vartheta+\pi\rangle\}$.
For angles $\vartheta=\pm 4\pi/3$ one finds equivalent pairs of
directional and planar orbitals in a 2D model,
$\{|i\zeta_a\rangle,|i\xi_a\rangle\}$ and
$\{|i\zeta_b\rangle,|i\xi_b\rangle\}$, to the usually used
~{$e_g$-orbital} real basis given by Eq.~(\ref{real}).

Consider now a bond $\langle ij\rangle\!\parallel\!\gamma$ along one of
the cubic axes $\gamma=a,b,c$, and a charge excitation generated by
a hopping process $i\rightarrow j$. The hopping $t$ couples two
directional orbitals
$\{|i\zeta_{\gamma}\rangle,|j\zeta_{\gamma}\rangle\}$.
Local projection operators on these active and the complementary
inactive $\{|i\xi_{\gamma}\rangle,|j\xi_{\gamma}\rangle\}$ orbitals
are
\beq
{\cal P}_{i\zeta}^{\gamma}=
|i\zeta_{\gamma}\rangle\langle i\zeta_{\gamma}|=
\left(\frac12+\tau^{(\gamma)}_i\right), \hskip .7cm
{\cal P}_{i\xi}^{\gamma}=
|i\xi_{\gamma}\rangle\langle i\xi_{\gamma}|=
\left(\frac12-\tau^{(\gamma)}_i\right),
\eeq
where
\beq
\tau^{(\gamma)}_i\equiv\,\frac12\,\left(
|i\zeta_{\gamma}\rangle\langle i\zeta_{\gamma}|-
|i  \xi_{\gamma}\rangle\langle i  \xi_{\gamma}|\right),
\eeq
and these operators are represented in the fixed
$\{|iz\rangle,|i\bar{z}\rangle\}$ basis as follows:
\beq
\label{orbop}
\tau^{(a)}_i= -\frac14\left(\sigma^z_i-\sqrt{3}\sigma^x_i\right), \hskip .7cm
\tau^{(b)}_i= -\frac14\left(\sigma^z_i+\sqrt{3}\sigma^x_i\right), \hskip .7cm
\tau^{(c)}_i= \frac12 \sigma^z_i.
\eeq
A charge excitation between two transition metal ions with partly filled
$e_g$-orbitals will arise by a hopping process between two active
orbitals, $|i\zeta_{\gamma}\rangle$ and $|j\zeta_{\gamma}\rangle$.
To capture such processes we introduce two projection operators
on the orbital states for each bond,
\begin{eqnarray}
\label{porbit}
{\cal P}_{\langle ij\rangle}^{(\gamma)}&\equiv&
\left(\frac12+\tau^{(\gamma)}_i\right)\left(\frac12-\tau^{(\gamma)}_j\right)+
\left(\frac12-\tau^{(\gamma)}_i\right)\left(\frac12+\tau^{(\gamma)}_j\right),  \\
\label{qorbit}
{\cal Q}_{\langle ij\rangle}^{(\gamma)}&\equiv&
2\left(\frac12-\tau^{(\gamma)}_i\right)\left(\frac12-\tau^{(\gamma)}_j\right).
\end{eqnarray}
Unlike for a spin system, the charge excitation
$d_i^md_j^m\rightleftharpoons d_i^{m+1}d_j^{m-1}$ is allowed only in
one direction when one orbital is directional $|\zeta_{\gamma}\rangle$
and the other is planar $|\xi_{\gamma}\rangle$ on the bond
$\langle ij\rangle\parallel\gamma$, i.e.,
$\left\langle{\cal P}_{\langle ij\rangle}^{(\gamma)}\right\rangle=1$;
such processes generate both HS and low-spin (LS) contributions. On the
contrary, when both orbitals are directional, i.e., one has
$\left\langle{\cal Q}_{\langle ij\rangle}^{(\gamma)}\right\rangle=2$,
only LS terms contribute.

\begin{figure}[t!]
\bec
\includegraphics[width=8cm,clip=true]{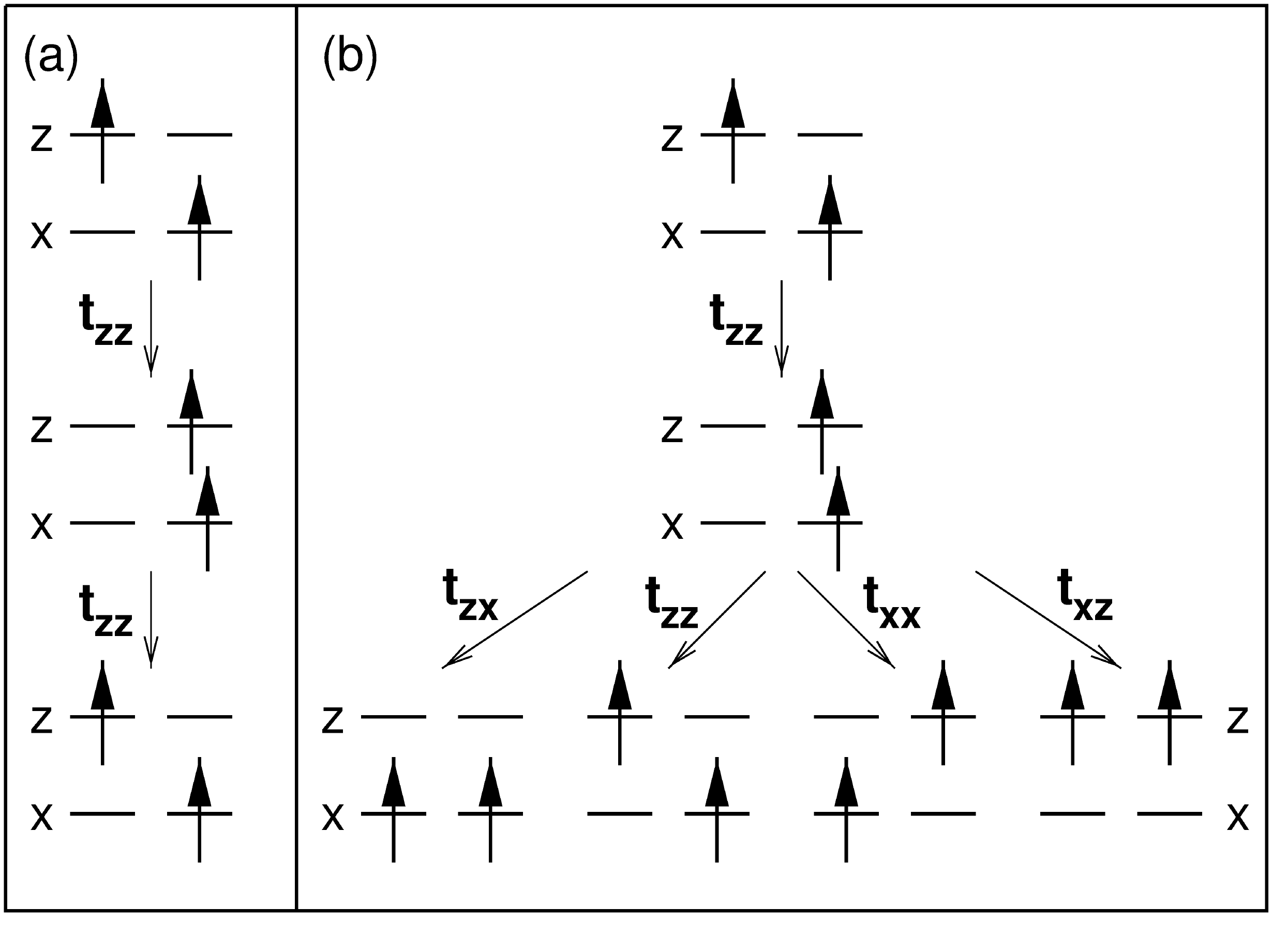}
\hskip .5cm
\includegraphics[width=7.1cm,clip=true]{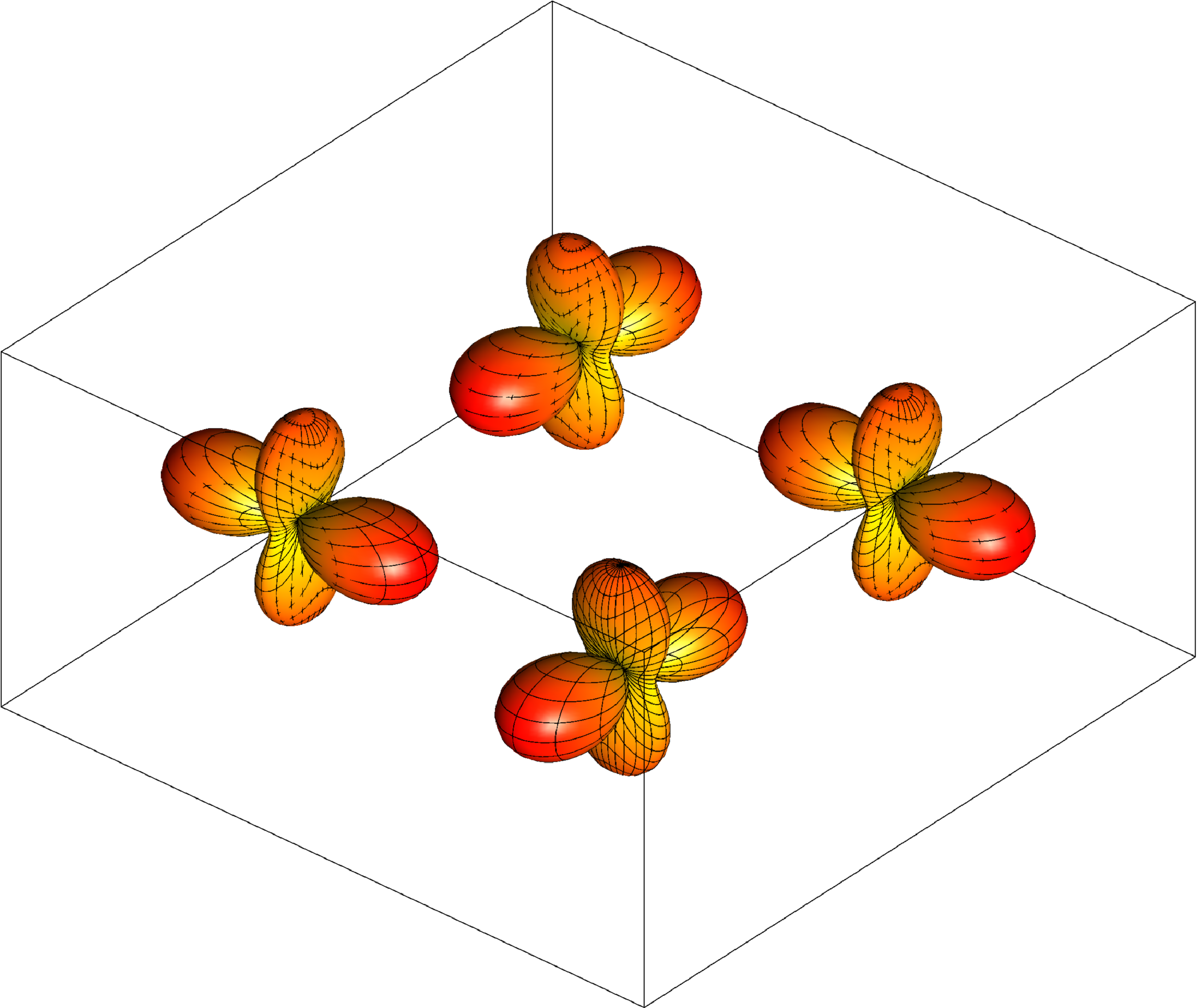} \hskip -.5cm (c)
\enc
\caption{Virtual charge excitations leading to the $e_g$ orbital
superexchange model for a strongly correlated system with $|z\rangle$ and
$|x\rangle\equiv|\bar{z}\rangle$ real $e_g$ orbitals
(\ref{real}) in the subspace of $\uparrow$-spin states:
(a)~for a bond along the $c$ axis $\langle ij\rangle\parallel c$;
(b)~for a bond  in the $ab$ plane $\langle ij\rangle\parallel ab$.
In a FM plane of KCuF$_3$ (LaMnO$_3$) the superexchange (\ref{osex})
favors (c) AO state of $|{\rm AO}\pm\rangle$ orbitals (\ref{pm}).
\hfill\break
Images (a-b) are reproduced from Ref. \cite{vdB99};
image (c) by courtesy of Krzysztof Bieniasz.
}
\label{fig:orbi}
\end{figure}

To write the superexchange model we need the charge excitation energy
which for the HS channel is,
\beq
\label{e1}
\varepsilon_1\equiv E_1(d^{m+1})+E_0(d^{m-1})-2E_0(d^m)=U-3J_H=\bar{U},
\eeq
where $E_0(d^m)$ in the ground state energy for an ion with $m$
electrons. Note that this energy is the same for KCuF$_3$ and LaMnO$_3$
\cite{Ole05}, so the $e_g$ orbital model presented here is universal.
Second order perturbation theory shown in Figs. \ref{fig:orbi}(a-b)
gives \cite{vdB99},
\beq
\label{egmo}
{\cal H}^{\uparrow}(e_g)=-\frac{t^2}{\varepsilon_1}
\sum_{\langle ij\rangle\parallel\gamma}
{\cal P}_{\langle ij\rangle}^{(\gamma)}.
\eeq
For convenience we define the dimensionless Hund's exchange parameter
$\eta$,
\bea
\label{eta}
\eta&\equiv&\frac{J_H}{U}.
\eea
The value of $J$ defines the superexchange energy scale and is the same
as in the $t$-$J$ model \cite{Cha77}, while the parameter $\eta$
(\ref{eta}) characterizes the multiplet structure when LS states
are included as well, see below. The $e_g$ orbital model (\ref{egmo})
(for HS states) takes the form,
\index{$e_g$ orbital superexchange!}
\beq
\label{osex}
{\cal H}^{\uparrow}(e_g)=\frac12 Jr_1 \sum_{\langle ij\rangle\parallel\gamma}
\left(\tau^{(\gamma)}_i\tau^{(\gamma)}_j-\frac14\right)
+E_z\sum_i\tau_i^{(c)},
\eeq
where $r_1=U/\varepsilon_{\rm HS}=U/\bar{U}=1/(1-3\eta)$. Here we
include the crystal-field term $\propto E_z$ which splits off the $e_g$
orbitals. The same effective model is obtained from the $e_g$ Hubbard
model Eq. (\ref{egHub}) at half-filling in the regime of $\bar{U}\gg t$.
It favors consistently with its derivation pairs of orthogonal orbitals
along the axis $\gamma$, with the energy gain for such a configuration
$-\frac14 Jr_1$. When both orbitals would be instead selected as
directional along the bond,
$\left\langle\tau^{(\gamma)}_i\tau^{(\gamma)}_j\right\rangle=\frac14$,
the energy gain vanishes as this orbital configuration corresponds
to the situation incompatible with the HS excited states considered
here and the superexchange is blocked. The ground state in the 2D $ab$
plane has alternating orbital (AO) order between the sublattices
$i\in A$ and $j\in B$,
\begin{equation}
\label{pm}
\textstyle{
|i+\rangle=\frac{1}{\sqrt{2}}\big(|iz\rangle +|i\bar{z}\rangle\big),
\hspace{0.7cm}
|j-\rangle=\frac{1}{\sqrt{2}}\big(|jz\rangle -|j\bar{z}\rangle\big),}
\end{equation}
of orbitals occupied by holes in KCuF$_3$ and by electrons in LaMnO$_3$,
see Fig. \ref{fig:orbi}(c).

\begin{figure}[t!]
\bec
\includegraphics[width=6.4cm,clip=true]{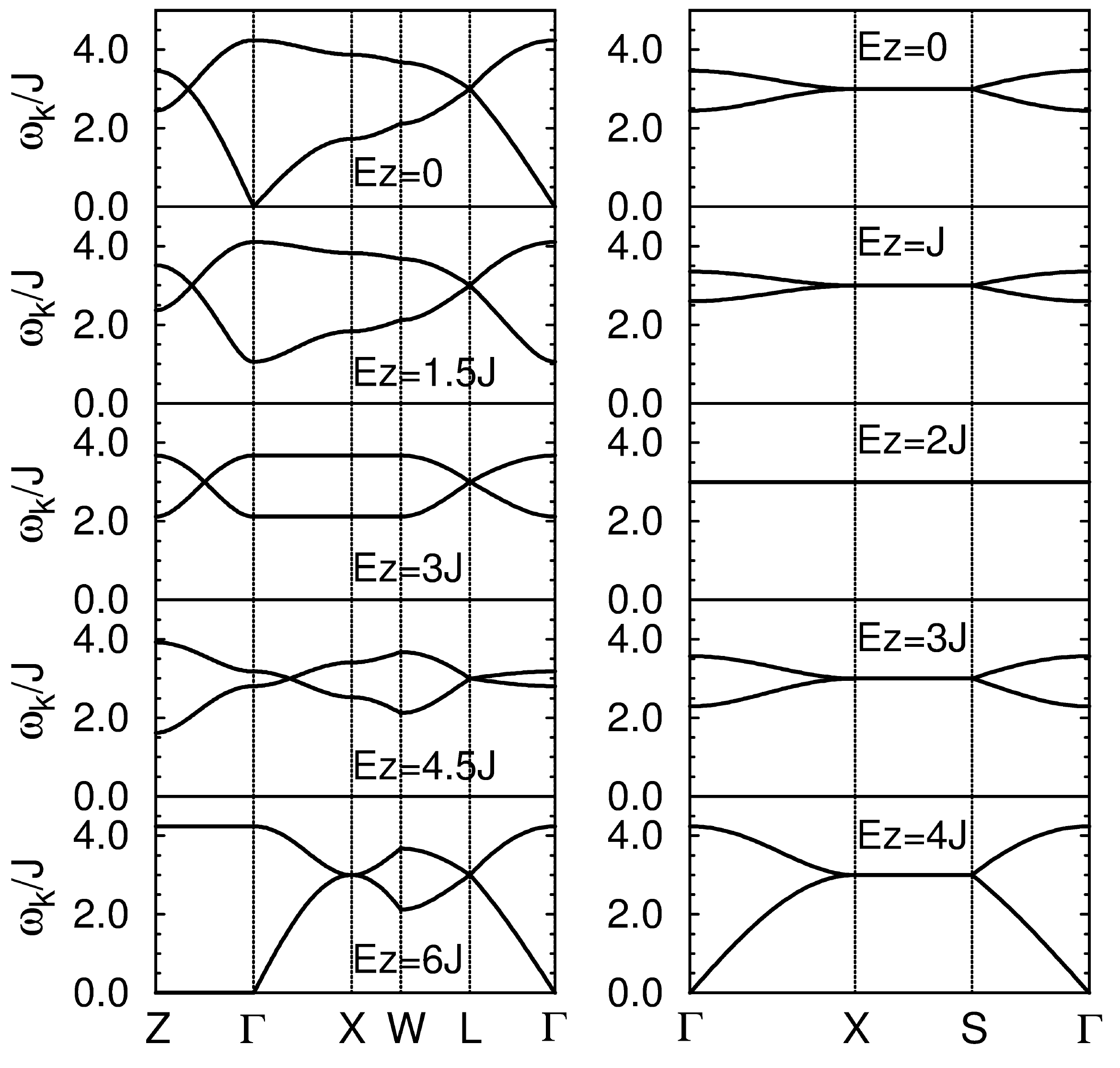} \hskip 0cm (a-b)
\hskip .1cm
\includegraphics[width=8.5cm,clip=true]{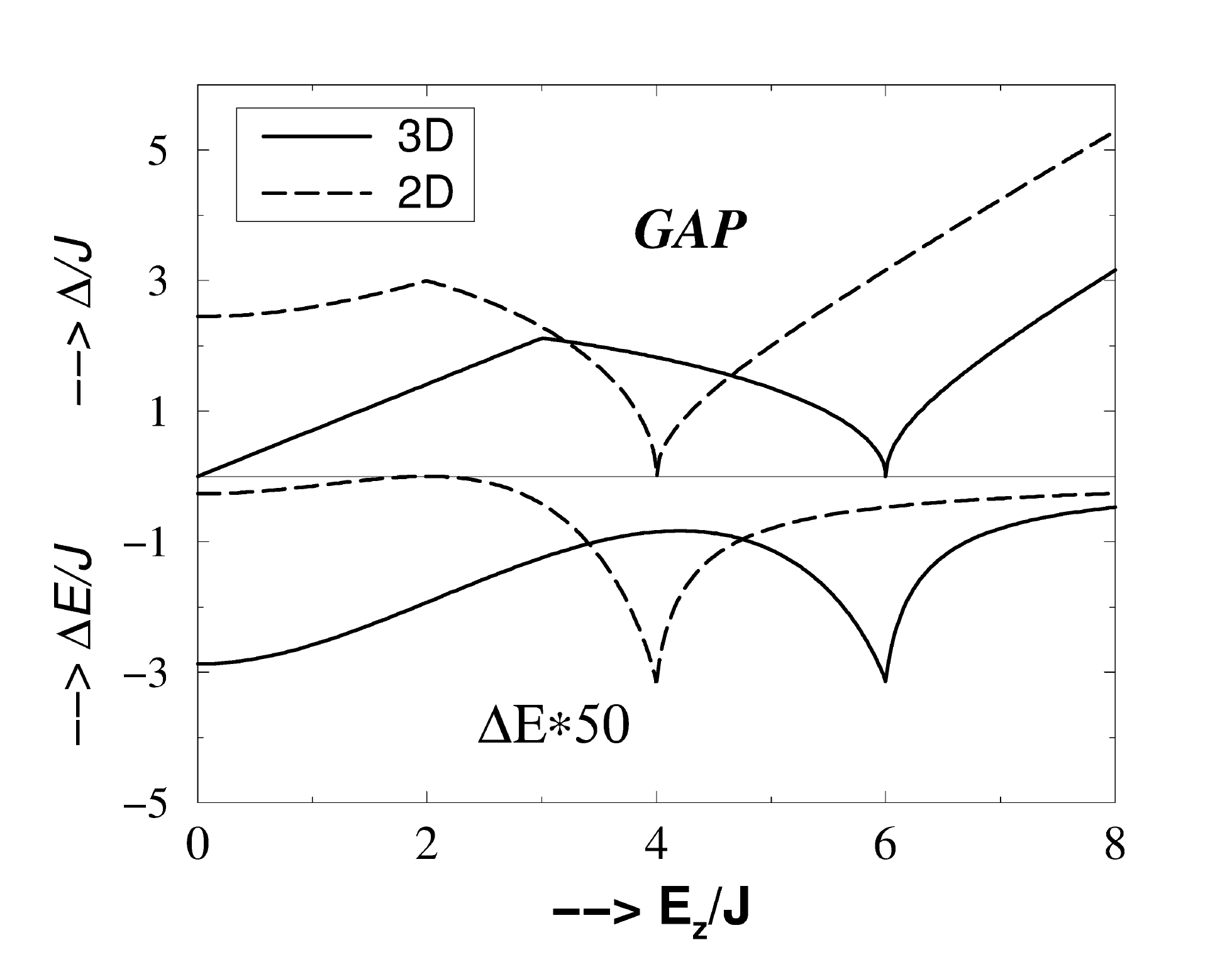} \hskip -1.2cm (c)
\enc
\caption{
(a-b) Orbital-wave excitations obtained for different values of the
crystal-field splitting $E_z$ for a 3D (left) and 2D (right) orbital
superexchange model (\ref{osex}), with $Jr_1\equiv J$. The result shown
for a 3D model at $E_z=0$ corresponds to the $E_z\to 0$ limit.
(c) Gap $\Delta/J$ in the orbital excitation spectrum and energy
quantum correction $\Delta E/J$ as functions of the crystal-field
splitting $E_z/J$, for the 3D (2D) model shown by full (dashed) lines.
\hfill\break
Images are reproduced from Ref. \cite{vdB99}.
}
\label{fig:owa}
\end{figure}

Here we are interested in the low temperature range $T<0.1J$ and the
2D (and 3D) $e_g$ orbital model orders at finite temperature $T<T_c$
\cite{Ryn10}, i.e., below $T_c=0.3566J$ for a 2D model \cite{Cza17},
so we assume perfect orbital order given by a classical \textit{Ansatz}
for the ground state,
\begin{equation}
\label{oo}
|\Phi_0\rangle=\prod_{i\in A}|i\theta_A\rangle
               \prod_{j\in B}|j\theta_B\rangle,
\end{equation}
with the orbital states, $|i\theta_A\rangle$ and $|j\theta_B\rangle$,
characterized by opposite angles ($\theta_A=-\theta_B$) and alternating
between two sublattices $A$ and $B$ in the $ab$ planes.
The orbital state at site $i$:
\begin{equation}
\label{mixing}
|i\theta\rangle=\cos\left(\theta/2\right)|iz\rangle
               +\sin\left(\theta/2\right)|i\bar{z}\rangle,
\end{equation}
is here parameterized by an angle $\theta$ which defines the
amplitudes of the orbital states defined in Eq. (\ref{real}).
The AO state specified in Eq. (\ref{oo}) is thus defined by:
\index{$e_g$ orbital superexchange!alternating order}
\begin{eqnarray}
\label{ood9}
|i\theta_A\rangle&=&\cos\left(\theta/2\right)|iz\rangle
                   +\sin\left(\theta/2\right)|ix\rangle,\nonumber\\
|j\theta_B\rangle&=&\cos\left(\theta/2\right)|jz\rangle
                   -\sin\left(\theta/2\right)|jx\rangle,
\end{eqnarray}
with $\theta_A=\theta$ and $\theta_B=-\theta$.

The excitations from the ground state of the orbital model (\ref{osex})
are orbital waves (orbitons) which may be obtained in a similar way to
magnons in a quantum antiferromagnet. An important difference is that
the orbitons have two branches which are in general nondegenerate, see
Fig. \ref{fig:owa}(a-b). In the absence of crystal field ($E_z=0$) the
spectrum for the 2D $e_g$ orbital model has a gap and the orbitons have
weak dispersion, so the quantum corrections to the order parameter
are rather small. They are much larger in the 3D model but still
smaller than in an antiferromagnet \cite{vdB99}. The gap closes in the
3D model at $E_z=0$, but the quantum corrections are smaller than in
the Heisenberg model. Note that the shape of the occupied orbitals
changes at finite crystal field, and the orbitons have a remarkable
evolution, both in the 3D and 2D model, see Figs. \ref{fig:owa}(a-b).
Increasing $E_z>0$ first increases the gap but when the field overcomes
the interactions and polarizes the orbitals (at $E_z=4J$ in 2D and
$E_z=6J$ in 3D model), the gap closes, see Fig. \ref{fig:owa}(c). This
point marks a transition from the AO order to uniform ferro-orbital (FO)
order. Note that in agreement with intuition the quantum corrections
$\Delta E/J$ are maximal when the gap closes and low-energy orbitons
contribute.
\index{$e_g$ orbital superexchange!orbitons}

\begin{figure}[t!]
\bec
\includegraphics[width=12cm,clip=true]{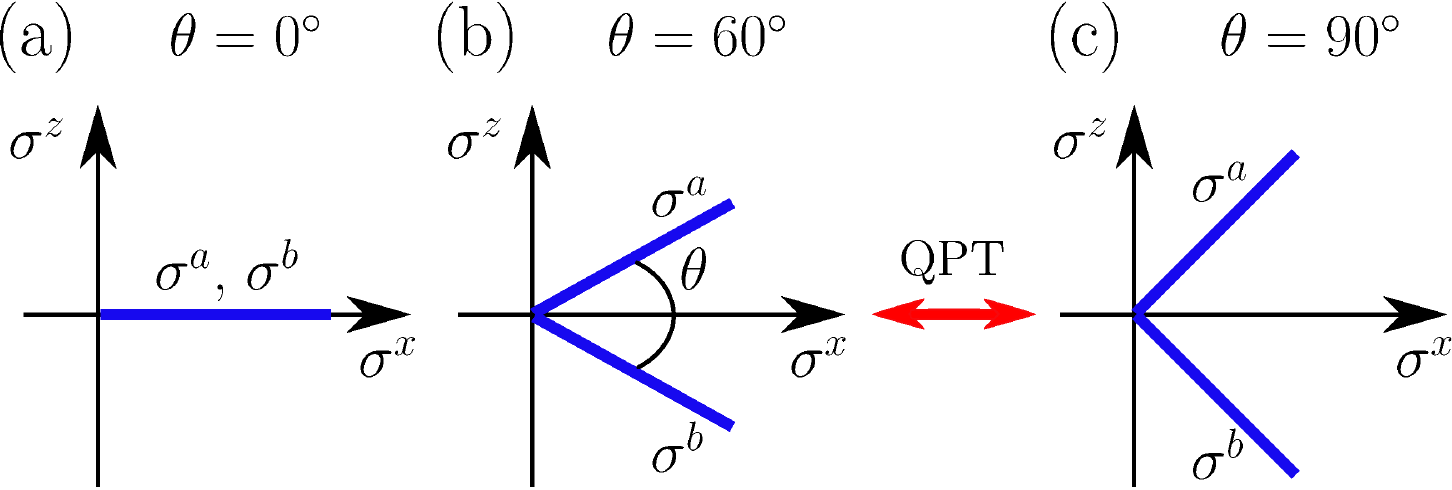}
\enc
\caption{Artist's view of the evolution of orbital interactions in the
generalized compass model Eq. (\ref{geno}) with increasing
angle $\theta$. Heavy (blue) lines indicate favored spin direction
induced by interactions along two nonequivalent lattice axes $a$ and
$b$. Different panels show: (a)~the Ising model at $\theta=0^\circ$,
(b)~the 2D $e_g$ orbital model at $\theta=60^\circ$, and
(c)~the OCM at $\theta=90^\circ$. Spin order follows the interactions
in the Ising limit, while it follows one of the equivalent interactions,
$\sigma^a$ or $\sigma^b$, in the OCM. This results in the symmetry
breaking quantum phase transition (QPT) which occurs between (b) and
(c). Image is reproduced from Ref. \cite{Cin10}.
}
\label{fig:ocm}
\end{figure}

To see the relation of the 2D $e_g$ orbital model to the compass model
\cite{vdB15} we introduce a 2D {\it generalized\/} compass model (GCM)
with pseudospin interactions on a square lattice in $ab$ plane
($J_{\rm cm}>0$) \cite{Cin10},
\index{compass model!generalized}
\begin{equation}
\label{geno}
{\cal H}(\theta)=-J_{\rm cm}\sum_{\{ij\}\in ab} \left\{
\sigma^a_{ij}(\theta)\sigma^a_{i+1,j}(\theta) +
\sigma^b_{ij}(\theta)\sigma^b_{i,j+1}(\theta) \right\}\,.
\end{equation}
The interactions occur along nearest neighbor bonds and are balanced
along both lattice directions $a$ and $b$. Here $\{ij\}$ labels lattice
sites in the $ab$ plane and
$\{\sigma^a_{ij}(\theta),\sigma^b_{ij}(\theta)\}$ are linear
combinations of Pauli matrices describing interactions for $T=\frac12$
pseudospins:
\begin{eqnarray}
\label{sigmaab}
\sigma^a_{ij}(\theta)\! &=& \cos(\theta/2)\;\sigma^x_{ij}
                           +\sin(\theta/2)\;\sigma^z_{ij}\,,\nn \\
\sigma^b_{ij}(\theta)\! &=& \cos(\theta/2)\;\sigma^x_{ij}
                           -\sin(\theta/2)\;\sigma^z_{ij}\,.
\end{eqnarray}
The interactions in Eq. (\ref{geno}) include the classical Ising model
for $\sigma^x_{ij}$ operators at $\theta=0^\circ$ and become gradually
more frustrated with increasing angle $\theta\in(0^\circ,90^\circ]$ ---
they interpolate between the Ising model (at $\theta=0^\circ$) and the
isotropic compass model (at $\theta=90^\circ$), see Fig.~\ref{fig:ocm}.
The latter case is equivalent by a standard unitary transformation to
the 2D compass model with standard interactions,
$\sigma^x_{ij}\sigma^x_{i,j+1}$ along the $a$ and
$\sigma^z_{ij}\sigma^z_{i+1,j}$ along the $b$ axis \cite{Cin10},
\index{compass model}
\begin{equation}
\label{com}
{\cal H}(\pi/2)=-J_{\rm cm}\sum_{\langle ij\rangle\parallel a}\,
\sigma^x_{ij}\sigma^x_{i+1,j}
-J_{\rm cm}\sum_{\langle ij\rangle\parallel b}\,
\sigma^z_{ij}\sigma^z_{i,j+1}  \,.
\end{equation}
The model (\ref{geno}) includes as well the 2D $e_g$ orbital model as a
special case, i.e., at $\theta=60^\circ$.
Increasing angle $\theta$ between the interacting orbital-like
components (\ref{sigmaab}) in Fig. \ref{fig:ocm} is equivalent to
increasing frustration which becomes maximal in the 2D compass model.
As a result, a second order quantum phase transition from Ising to
nematic order \cite{Wen11} occurs at $\theta_c\simeq 84.8^\circ$ which
is surprisingly close to the compass point $\theta=90^\circ$, i.e., only
when the interactions are sufficiently strongly frustrated. The ground
state has high degeneracy $d=2^{L+1}$ for a 2D cluster $L\times L$ of
one-dimensional (1D) nematic states which are entirely different from
the 2D AO order in the $e_g$ orbital model depicted in
Fig.~\ref{fig:ocm}(c), yet it is stable in a range of temperature below
$T_c\simeq 0.06J_{\rm cm}$ \cite{Cza16}.



\section{Superexchange models for active $e_g$ orbitals}
\label{sexe}

\subsection{General structure of the spin-orbital superexchange}
\label{general}

\begin{figure}[t!]
\bec
\includegraphics[width=12cm,clip=true]{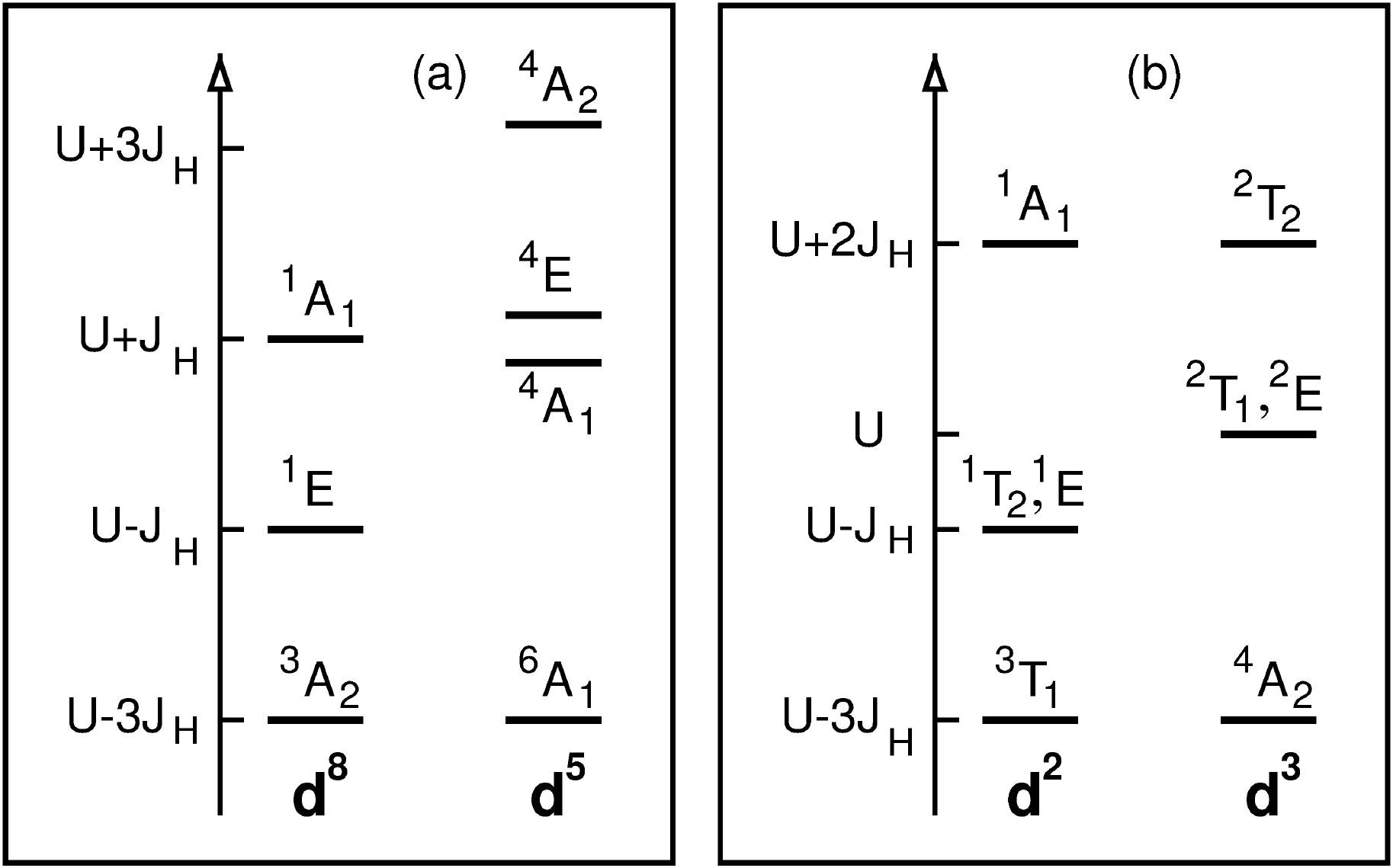}
\enc
\caption{Energies of charge excitations $\varepsilon_n$ (\ref{excen})
for selected cubic transition metal oxides, for:
(a) $e_g$ excitations to Cu$^{3+}$ ($d^8$) and Mn$^{2+}$ ($d^5$) ions;
(b) $t_{2g}$ excitations to Ti$^{2+}$ ($d^2$) and V$^{2+}$ ($d^3$)
ions.
The splittings between different states are due to Hund's exchange
element $J_H$ which refers to a pair of $e_g$ and $t_{2g}$ electrons
in (a) and (b).
Image is reproduced from Ref. \cite{Ole05}.
}
\label{fig:exci}
\end{figure}

We consider the case of partly filled degenerate $3d$-orbitals and large
Hund's exchange $J_H$. In the regime of $t\ll U$, electrons localize
and effective low-energy superexchange interactions consist of all the
contributions which originate from possible virtual charge excitations,
$d_i^md_j^m\rightleftharpoons d_i^{m+1}d_j^{m-1}$ --- they take a form
of a spin-orbital model, see Eq. (\ref{som}) below.
The charge excitation $n$ costs the energy
\beq
\label{excen}
\varepsilon_n=E_n(d^{m+1})+E_0(d^{m-1})-2E_0(d^m),
\eeq
where the $d^m$ ions are in the initial HS ground states with spins
$S=\frac{m}{2}$ and have the Coulomb interaction energy
$E_0(d^m)={m\choose 2}(U-3J_H)$ each (if $m<5$, else if $m>5$ one has
to consider here $m$ holes instead, while the case of $m=5$ is special
and will not be considered here as in the $t_{2g}^3e_g^2$ configuration
the orbital degree of freedom is quenched). The same formula for ground
state energy applies as well to Mn$^{3+}$ ions in $d^4$ configuration
with $S=2$ spin HS ground state, see Sec. \ref{lamno}. By construction
also the ion with less electrons (holes) for $m<5$ is in the HS state
and $E_0(d^{m-1})={m-1\choose 2}(U-3J_H)$. The excitation energies
(\ref{excen}) are thus defined by the multiplet structure of an ion
with more electrons (holes) in the configuration $d^{m+1}$, see
Fig. \ref{fig:exci}. The lowest energy excitation is given by
Eq. (\ref{e1}) --- it is obtained from the HS state of the $3d^{m+1}$
ion with total spin
${\cal S}=S+\frac12$ and energy $E_1(d^{m+1})={m+1\choose 2}(U-3J_H)$.
Indeed, one recovers the lowest
excitation energy in the HS subspace, see Eq.~(\ref{e1}), with $J_H$
being Hund's exchange element for the electron (hole) involved in the
charge excitation, either $e_g$ or $t_{2g}$. We emphasize that this
lowest excitation energy $\varepsilon_1$ (\ref{e1}) is universal and
is found both in $t_{2g}$ and $e_g$ systems, i.e., it does not depend
on the electron valence $m$. In contrast, the remaining energies
$\{\varepsilon_n\}$ for $n>1$ are all for LS excitations and are
specific to a given valence $m$ of the considered insulator with $d^m$
ions. They have to be determined from the full local Coulomb
interaction Hamiltonian (\ref{Hub}), in general including also the
anisotropy of $\{U_{\alpha\beta}\}$ and $\{J_{\alpha\beta}\}$ elements.

Effective interactions in a Mott (or charge transfer) insulator with
orbital degeneracy take the form of spin-orbital superexchange
\cite{Kug82,Kha05}.
Its general structure is given by the sum over all the nearest neighbor
bonds $\langle ij\rangle\!\parallel\!\gamma$ connecting two transition
metal ions and over the excitations $n$ possible for each of them as,
\index{Mott insulator}
\beq
\label{Hsex}
{\cal H}=-\sum_n\frac{t^2}{\varepsilon_n}
\sum_{\langle ij\rangle\parallel\gamma}\,
P_{\langle ij\rangle}({\cal S})\,{\cal O}_{\langle ij\rangle}^{\gamma},
\eeq
where $P_{\langle ij\rangle}({\cal S})$ is the projection on the total
spin ${\cal S}=S\pm\frac12$ and ${\cal O}_{\langle ij\rangle}^{\gamma}$
is the projection operator on the orbital state at the sites $i$ and $j$
of the bond. Following this general procedure, one finds a spin-orbital
model with Heisenberg spin interaction for spins $S=\frac{m}{2}$ of
SU(2) symmetry coupled to the orbital operators which have much lower
cubic symmetry, with the general structure of spin-orbital
superexchange $\propto J$ (\ref{J}) \cite{Ole05},
\index{spin-orbital superexchange}
\begin{equation}
\label{som}
{\cal H}_J = J \sum_{\gamma}\sum_{\langle ij \rangle\parallel\gamma} \left\{
{\hat {\cal K}}_{ij}^{(\gamma)}
\left( {\vec S}_i \cdot {\vec S}_j +S^2\right) +
{\hat {\cal N}}_{ij}^{(\gamma)} \right\}.
\end{equation}
It connects ions at sites $i$ and $j$ along the bond
$\langle ij\rangle\parallel\gamma$ and involves orbital operators,
${\hat{\cal K}}_{ij}^{(\gamma)}$ and ${\hat{\cal N}}_{ij}^{(\gamma)}$
which depend on the bond direction $\gamma=a,b,c$ for the three
{\it a priori\/} equivalent directions in a cubic crystal. The spin
scalar product, $\left({\vec S}_i\cdot{\vec S}_j\right)$, is coupled to
orbital operators ${\hat{\cal K}}_{ij}^{(\gamma)}$ which together with
the other "decoupled" orbital operators, ${\hat{\cal N}}_{ij}^{(\gamma)}$,
determine the orbital state in a Mott insulator. The form of these
operators depends on the type of orbital degrees of freedom
in a given model. They involve active orbitals on each bond
$\langle ij \rangle\parallel\gamma$ along direction $\gamma$. Thus the
orbital interactions are directional and have only the cubic symmetry
of a (perovskite) lattice provided the symmetry in the orbital sector
is not broken by other interactions, for instance by crystal-field or
Jahn-Teller terms.

The magnetic superexchange constants along each cubic axis $J_{ab}$ and
$J_{c}$ in the effective spin model,
\beq
\label{Hs}
H=J_{ab}\sum_{\langle ij\rangle\parallel ab}{\vec S}_i\cdot{\vec S}_j
 +J_c   \sum_{\langle ij\rangle\parallel  c}{\vec S}_i\cdot{\vec S}_j,
\eeq
are obtained from the spin-orbital model (\ref{som}) by decoupling spin
and orbital operators and next averaging the orbital operators over a
given orbital (ordered or disordered) state. It gives effective
magnetic exchange interactions: $J_c$ along the $c$ axis, and $J_{ab}$
within the $ab$ planes. The latter $J_{ab}$ ones could in principle
still be different between the $a$ and $b$ axes in case of finite
lattice distortions due to the Jahn-Teller effect or octahedra tilting,
but we limit ourselves to idealized structures with $J_{ab}$ being the
same for both planar directions. We show below that the spin-spin
correlations along the $c$ axis and within the $ab$ planes,
\begin{equation}
\label{spins}
   s_c=\langle\vec{S}_i\cdot\vec{S}_j\rangle_{c},   \hskip .7cm
s_{ab}=\langle\vec{S}_i\cdot\vec{S}_j\rangle_{ab},
\end{equation}
next to the orbital correlations, play an important role in the
intensity distribution in optical spectroscopy.

In the correlated insulators with partly occupied degenerate orbitals
not only the structure of the superexchange (\ref{som}) is complex, but
also the optical spectra exhibit strong anisotropy and temperature
dependence near the magnetic transitions, as found e.g. in LaMnO$_3$
\cite{Kov10} or in the cubic vanadates LaVO$_3$ and YVO$_3$
\cite{Miy02}. In such systems several excitations contribute to the
excitation spectra, so one may ask how the spectral weight
redistributes between individual subbands originating from these
excitations. The spectral weight distribution is in general anisotropic
already when orbital order sets in and breaks the cubic symmetry, but
even more so when $A$-type or $C$-type AF spin order occurs below the
N\'eel temperature $T_{\rm N}$.

At orbital degeneracy the superexchange consists of the terms
$H_n^{(\gamma)}(ij)$ as a superposition of individual contributions on
each bond $\langle ij\rangle$ due to charge excitation $n$
(\ref{excen}) \cite{Kha04},
\begin{equation}
\label{HJ}
{\cal H} = J \sum_n\sum_{\langle ij\rangle\parallel\gamma}
             H_n^{(\gamma)}(ij),
\end{equation}
with the energy unit for each individual $H_n^{(\gamma)}(ij)$ term
given by the superexchange constant $J$ (\ref{J}). It follows from $d-d$
charge excitations with an effective hopping element $t$ between
neighboring transition metal ions and is the same as that obtained in a
Mott insulator with nondegenerate orbitals in the regime of $U\gg t$.
The spectral weight in the optical spectroscopy is determined by the
kinetic energy, and reflects the onset of magnetic order and/or orbital
order \cite{Kha04}. In a correlated insulator the electrons are almost
localized and the only kinetic energy which is left is associated with
the same virtual charge excitations that contribute also to the
superexchange. Therefore, the individual kinetic energy terms
$K_n^{(\gamma)}$ may be directly determined from the superexchange
(\ref{HJ}) using the Hellman-Feynman theorem,
\index{optical spectral weight}
\begin{equation}
\label{hefa}
K_n^{(\gamma)}=-2J\left\langle H_n^{(\gamma)}(ij)\right\rangle.
\end{equation}
For convenience, we define here the $K_n^{(\gamma)}$ as positive
quantities. Each term $K_n^{(\gamma)}$ (\ref{hefa}) originates from a
given charge excitation $n$ along a bond
$\langle ij\rangle\parallel\gamma$.
These terms are directly related to the {\it partial optical sum rule\/}
for individual Hubbard subbands, which reads \cite{Kha04}
\begin{equation}
\label{opsa}
\frac{a_0\hbar^2}{e^2}\int_0^{\infty}\sigma_n^{(\gamma)}(\omega)d\omega=
\frac{\pi}{2}K_n^{(\gamma)},
\end{equation}
where $\sigma_n^{(\gamma)}(\omega)$ is the contribution of band $n$ to
the optical conductivity for polarization along the $\gamma$ axis,
$a_0$ is the distance between transition metal ions, and the
tight-binding model with nearest neighbor hopping is implied.
Using Eq. (\ref{hefa}) one finds that the intensity of each band is
indeed determined by the underlying orbital order together with the
spin-spin correlation along the direction corresponding to the
polarization.

One has to distinguish the above partial sum rule (\ref{opsa}) from the
full sum rule for the total spectral weight in the optical spectroscopy
for polarization along a cubic direction $\gamma$, involving
\begin{equation}
\label{opsatot}
K^{(\gamma)}=-2J\sum_n\big\langle H_n^{(\gamma)}(ij)\big\rangle,
\end{equation}
which stands for the total intensity in the optical $d-d$ excitations.
This quantity is usually of less interest as it does not allow for a
direct insight into the nature of the electronic structure being a sum
over several excitations with different energies $\varepsilon_n$
(\ref{excen}) and has a much weaker temperature dependence.
In addition, it might be also more difficult to deduce from experiment.

\subsection{Kugel-Khomskii model for KCuF$_3$ and K$_2$CuF$_4$}
\label{sec:KK}

The simplest and seminal spin-orbital model is obtained when a fermion
has two flavors, spin and orbital, and both have two components, i.e.,
spin and pseudospin are $S=T=\frac12$. The physical realization is
found in cuprates with degenerate $e_g$ orbitals, such as KCuF$_3$ or
K$_2$CuF$_4$ \cite{Kug82}, where Cu$^{2+}$ ions are in the $d^9$
electronic configuration, so charge excitations
$d_i^9d_j^9\rightleftharpoons d_i^{10}d_j^8$ are made by holes.
By considering the degenerate Hubbard model for two $e_g$ orbitals one
finds that $d^8$ ions have an equidistant multiplet structure, with
three excitation energies which differ by $2J_H$ [here $J_H$ stands for
$J_H^e$ in Eq. (\ref{JHe})], see Table~2. We emphasize that the correct
spectrum has a doubly degenerate energy $(U-J_H)$ and the highest
non-degenerate energy is $(U+J_H)$, see Fig. \ref{fig:exci}(a).
Note that this result follows from the diagonalization of the local
Coulomb interactions in the relevant subspaces --- it reflects the fact
that a double occupancy ($|z{\uparrow}z{\downarrow}\rangle$ or
$|\bar{z}{\uparrow}\bar{z}{\downarrow}\rangle$) in either orbital state
($|z\rangle$ or $|\bar{z}\rangle$) is not an eigenstate of the
degenerate Hubbard in the atomic limit (\ref{Hee}), so the excitation
energy $U$ is absent in the spectrum, see Table~2.

\begin{table}[b!]
\caption{
Elements needed for the construction of the Kugel-Khomskii model from
charge excitations on the bond $\langle ij\rangle$:
excitation $n$, its type (HS or LS) and energy $\varepsilon_n$, total
spin state (triplet or singlet) and the spin projection operator
$P_{\langle ij\rangle}({\cal S})$,
and the orbital state and the corresponding orbital projection operator. }
\vskip .2cm
\bec
\begin{tabular}{ccccccc}
\hline\hline
 \multicolumn{3}{c}{charge excitation} & \multicolumn{2}{c}{spin state}
                                       & \multicolumn{2}{c}{orbital state} \cr
  $n$ & type & $\varepsilon_n$ &  ${\cal S}$ & $P_{\langle ij\rangle}({\cal S})$
             & orbitals on $\langle ij\rangle\parallel\gamma$ & projection \cr
\hline
   1  &  HS  &    $U-3J_H$     & $1$ &
$\,\,\,\,\left({\vec S}_i\cdot{\vec S}_j+\frac34\right)$ &
$|i\zeta_\gamma\rangle\,|j\xi_\gamma\rangle$
$\left(|i\xi_\gamma\rangle\,|j\zeta_\gamma\rangle\right)$
      & ${\cal P}_{\langle ij\rangle}^{(\gamma)}$ \cr
   2  &  LS  &    $U- J_H$     & $0$ &
$-\left({\vec S}_i\cdot{\vec S}_j-\frac14\right)$ &
$|i\zeta_\gamma\rangle\,|j\xi_\gamma\rangle$
$\left(|i\xi_\gamma\rangle\,|j\zeta_\gamma\rangle\right)$
      & ${\cal P}_{\langle ij\rangle}^{(\gamma)}$ \cr
   3  &  LS  &    $U- J_H$     & $0$ &
$-\left({\vec S}_i\cdot{\vec S}_j-\frac14\right)$ &
$|i\zeta_\gamma\rangle\,|j\zeta_\gamma\rangle$
      & ${\cal Q}_{\langle ij\rangle}^{(\gamma)}$ \cr
   4  &  LS  &    $U+ J_H$     & $0$ &
$-\left({\vec S}_i\cdot{\vec S}_j-\frac14\right)$ &
$|i\zeta_\gamma\rangle\,|j\zeta_\gamma\rangle$
      & ${\cal Q}_{\langle ij\rangle}^{(\gamma)}$ \cr
\hline\hline
\end{tabular}
\enc
\label{tab:kk}
\end{table}

The total spin state on the bond corresponds to ${\cal S}=1$ or 0,
so the spin projection operators $P_{\langle ij\rangle}(1)$ and
$P_{\langle ij\rangle}(0)$ are easily deduced, see Table~2. The
orbital configuration which corresponds to a given bond
$\langle ij\rangle$ is given by one of the orbital operators in
Sec. \ref{sec:orbi}, either ${\cal P}_{\langle ij\rangle}^{(\gamma)}$
for the doubly occupied states involving different orbitals, or
${\cal Q}_{\langle ij\rangle}^{(\gamma)}$ for a double occupancy in
a directional orbital at site $i$ or $j$. This gives a rather
transparent structure of one HS and three LS excitations in Table~2.
The 3D Kugel-Khomskii (KK) model then follows from Eq.~(\ref{Hsex})
\cite{Fei97,Ole00}:
\bea
\label{d9}
{\cal H}(d^9)&=&\sum_{\gamma}\sum_{\langle ij\rangle\parallel\gamma}
\left\{-\frac{t^2}{U-3J_H}\left({\vec S}_i\cdot{\vec S}_j+\frac34\right)
{\cal P}_{\langle ij\rangle}^{(\gamma)}
+\frac{t^2}{U- J_H}\left({\vec S}_i\cdot{\vec S}_j-\frac14\right)
{\cal P}_{\langle ij\rangle}^{(\gamma)}\right.\nn \\
& &\hskip 1.8cm \left.+\left(\frac{t^2}{U- J_H}+\frac{t^2}{U+ J_H}\right)
\left({\vec S}_i\cdot{\vec S}_j-\frac14\right)
{\cal Q}_{\langle ij\rangle}^{(\gamma)}\right\}+E_z\sum_i\tau_i^c.
\eea
The last term $\propto E_z$ is the crystal field which splits off the
degenerate $e_g$ orbitals when Jahn-Teller lattice distortion occurs,
and is together with Hund's exchange $\eta$ a second parameter to
construct phase diagrams, see below.
Here it refers to holes, i.e., large $E_z>0$ favors hole occupation in
$|\bar{z}\rangle\equiv|x^2-y^2\rangle/\sqrt{2}$ orbitals, as in
La$_2$CuO$_4$. On the other hand, while $E_z\simeq 0$, both orbitals
have almost equal hole density.

Another form of the
Hamiltonian (\ref{d9}) is obtained by introducing the coefficients,
\beq
\label{ri}
r_1=\frac{1}{1-3\eta},    \hskip .7cm
r_2=r_3=\frac{1}{1-\eta}, \hskip .7cm
r_4=\frac{1}{1+\eta},
\eeq
and defining the superexchange constant $J$ in the same way as in the
$t-J$ model Eq. (\ref{J}).
With the explicit representation of the orbital operators
${\cal P}_{\langle ij\rangle}^{(\gamma)}$ and
${\cal Q}_{\langle ij\rangle}^{(\gamma)}$
in terms of $\left\{\tau_i^{(\gamma)}\right\}$ one finds,
\index{Kugel-Khomskii model}
\index{spin-orbital superexchange!for KCuF$_3$ }
\bea
\label{KK}
{\cal H}(d^9)\!\!&=&\!\frac12 J\sum_{\gamma}\sum_{\langle ij\rangle\parallel\gamma}
\left\{\left[-r_1\left({\vec S}_i\cdot{\vec S}_j+\frac34\right)
+r_2\left({\vec S}_i\cdot{\vec S}_j-\frac14\right)\right]
\left(\frac14-\tau_i^{(\gamma)}\tau_j^{(\gamma)}\right)\right.\nn \\
\!\!& &\!\hskip .7cm \left.+\left(r_3+r_4\right)
\left({\vec S}_i\cdot{\vec S}_j-\frac14\right)
\left(\tau_i^{(\gamma)}+\frac12\right)\left(\tau_j^{(\gamma)}+\frac12\right)\right\}
+E_z\sum_i\tau_i^c.
\eea
In the FM state spins are integrated out and one finds from the first
term just the superexchange in the $e_g$ orbital model analyzed before
in Sec.~\ref{sec:orbi}.

The magnetic superexchange constants $J_{ab}$ and $J_{c}$ in the
effective spin-orbital model (\ref{KK}) are obtained by decoupling spin
and orbital operators and next averaging the orbital operators
$\left\langle{\hat {\cal K}}_{ij}^{(\gamma)}\right\rangle$ over the
classical state $|\Phi_0\rangle$ as given by Eq. (\ref{oo}). The
relevant averages are given in Table \ref{tab:eg}, and they lead to the
following expressions for the superexchange constants in Eq. (\ref{Hs}),
\begin{eqnarray}
\label{jc9}
J_c &=&\frac{1}{8}J\Big\{-r_1\sin^2\theta+(r_2+r_3)(1+\cos\theta)
+r_4(1+\cos\theta)^2\Big\},                  \\
\label{jab9}
J_{ab}&=&\frac{1}{8}J\left\{-r_1\left(\frac{3}{4}+\sin^2\theta\right)
    +(r_2+r_3)\left(1-\frac{1}{2}\cos\theta\right)
    + r_4\left(\frac{1}{2}-\cos\theta\right)^2\right\},
\end{eqnarray}
which depend on two parameters: $J$ (\ref{J}) and $\eta$ (\ref{eta}),
and on the orbital order (\ref{ood9}) specified by the orbital angle
$\theta$. It is clear that the FM term $\propto r_1$ competes with all
the other AF LS terms. Nevertheless, in the $ab$ planes, where the occupied
hole $e_g$ orbitals alternate, the larger FM contribution dominates and
makes the magnetic superexchange $J_{ab}$ weakly FM ($J_{ab}\lesssim 0$)
(when $\sin^2\theta\simeq 1$), while the stronger AF superexchange
along the $c$ axis ($J_{c}\gg |J_{ab}|$) favors quasi one-dimensional
(1D) spin fluctuations. Thus KCuF$_3$ exhibits spinon excitations for
$T>T{\rm N}$.

\begin{figure}[t!]
\bec
\includegraphics[width=7.2cm,clip=true]{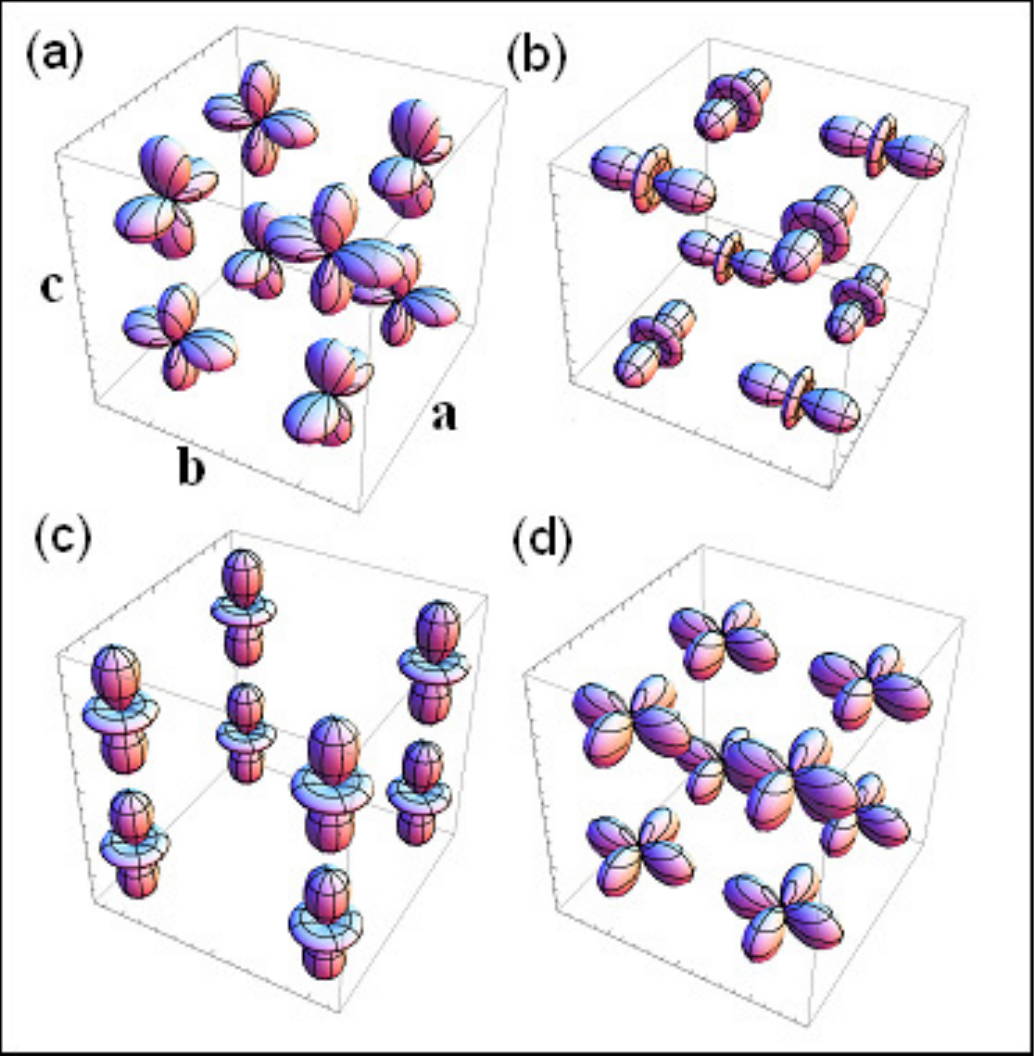}
\hskip .7cm
\includegraphics[width=7.7cm,clip=true]{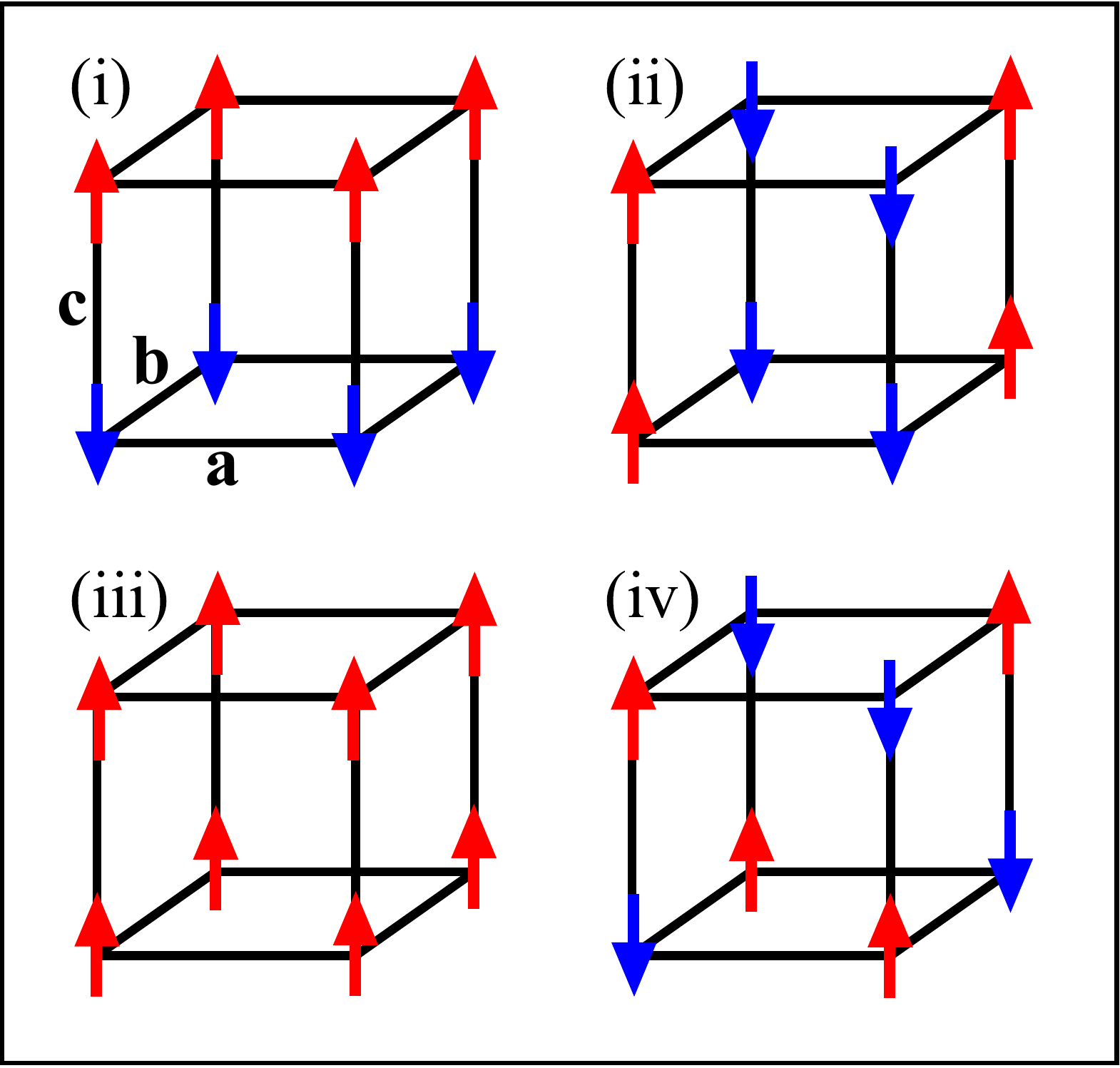}
\enc
\caption{
Left --- schematic view of four simplest orbital configurations
on a representative cube of the 3D lattice:
(a) AO order with $\langle\tau_{i}^{a(b)}\rangle=\pm\frac12$ changing
from site to site and $\langle\tau_{i}^{c}\rangle=\frac14$,
obtained for $E_{z}<0$,
(b) AO order with $\langle\tau_{i}^{a(b)}\rangle=-\frac12$ changing
from site to site and $\langle\tau_{i}^{c}\rangle=-\frac14$,
obtained for $E_{z}>0$,
(c) FO order with occupied $z$ orbitals and
$\langle\tau_{i}^{c}\rangle=\frac12$ (cigar-shaped orbitals), and
(d) FO order with occupied $\bar{z}$ orbitals and
$\langle\tau_{i}^{c}\rangle=-\frac12$ (clover-shaped orbitals).
Right --- schematic view of four spin configurations
(arrows stand for up or down spins) in phases with spin order:
(i) $A$-AF,
(ii) $C$-AF,
(iii) FM, and
(iv) $G$-AF.
Images are reproduced from Ref. \cite{Brz13}.}
\label{fig:so}
\end{figure}

\begin{table}[b!]
\caption{
Averages of the orbital projection operators standing in the
spin-orbital interactions in the KK model (\ref{KK}) and determine the
spin interactions in $H_s$ (\ref{Hs}) for the
$C$-type orbital order of occupied $e_g$ orbitals which alternate in
$ab$ planes, as given by Eqs. (\ref{ood9}). Nonequivalent cubic
directions are labeled by $\gamma=ab,c$.
}
\bec
\begin{tabular}{cccccc}\hline\hline
operator & average && $ab$ &&  $c$ \cr
\hline
${\cal Q}_{\langle ij\rangle}^{(\gamma)}$ &
$2\left\langle\big(\frac{1}{2}-\tau_i^{(\gamma)}\big)
            \big(\frac{1}{2}-\tau_j^{(\gamma)}\big)\right\rangle$ &&
	$\frac12\big(\frac{1}{2}-\cos\theta\big)^2$         &&
	$\frac12\big(1+\cos\theta\big)^2$                   \cr
${\cal P}_{\langle ij\rangle}^{(\gamma)}$ &
$\left\langle\frac{1}{4}-\tau_i^{(\gamma)}\tau_j^{(\gamma)}\right\rangle$ &&
        $\frac{1}{4}\big(\frac{3}{4}+\sin^2\theta\big)$         &&
	$\frac{1}{4}\sin^2\theta$ \cr
${\cal R}_{\langle ij\rangle}^{(\gamma)}$ &
$2\left\langle\big(\frac{1}{2}+\tau_i^{(\gamma)}\big)
            \big(\frac{1}{2}+\tau_j^{(\gamma)}\big)\right\rangle$ &&
	$\frac12\big(\frac{1}{2}+\cos\theta\big)^2$         &&
	$\frac12\big(1-\cos\theta\big)^2$                   \cr
\hline\hline
\end{tabular}
\enc
\label{tab:eg}
\end{table}

Consider first the 2D KK model on a square lattice, with $\gamma=a,b$
in Eq. (\ref{KK}), as in K$_2$CuF$_4$. In the absence of Hund's
exchange, interactions between $S=\frac12$ spins are AF. However, they
are quite different depending on which of the two $e_g$ orbitals are
occupied by holes: $J_{ab}^z=\frac{1}{16}J$ for $|z\rangle$ and
$J_{ab}^{\bar{z}}=\frac{9}{16}J$ for $|\bar{z}\rangle$ hole orbitals.
As a result, the AF phases with spin order in Fig. \ref{fig:so}(iv) and
the FO order shown in Figs.~\ref{fig:so}(c) and \ref{fig:so}(d) are
degenerate at finite crystal field $E_z=-\frac12 J$. This defines a
quantum critical point $Q_{2{\rm D}}=(-0.5,0)$ in the $(E_z/J,\eta)$
plane. Actually, at this point also one more phase has the same energy
--- the FM spin phase of Fig. \ref{fig:so}(i)
with AO order of $|\pm\rangle$ orbitals (\ref{pm}) shown in
Fig. \ref{fig:so}(a) \cite{Ole00}.

\begin{figure}[t!]
\bec
\includegraphics[width=15cm]{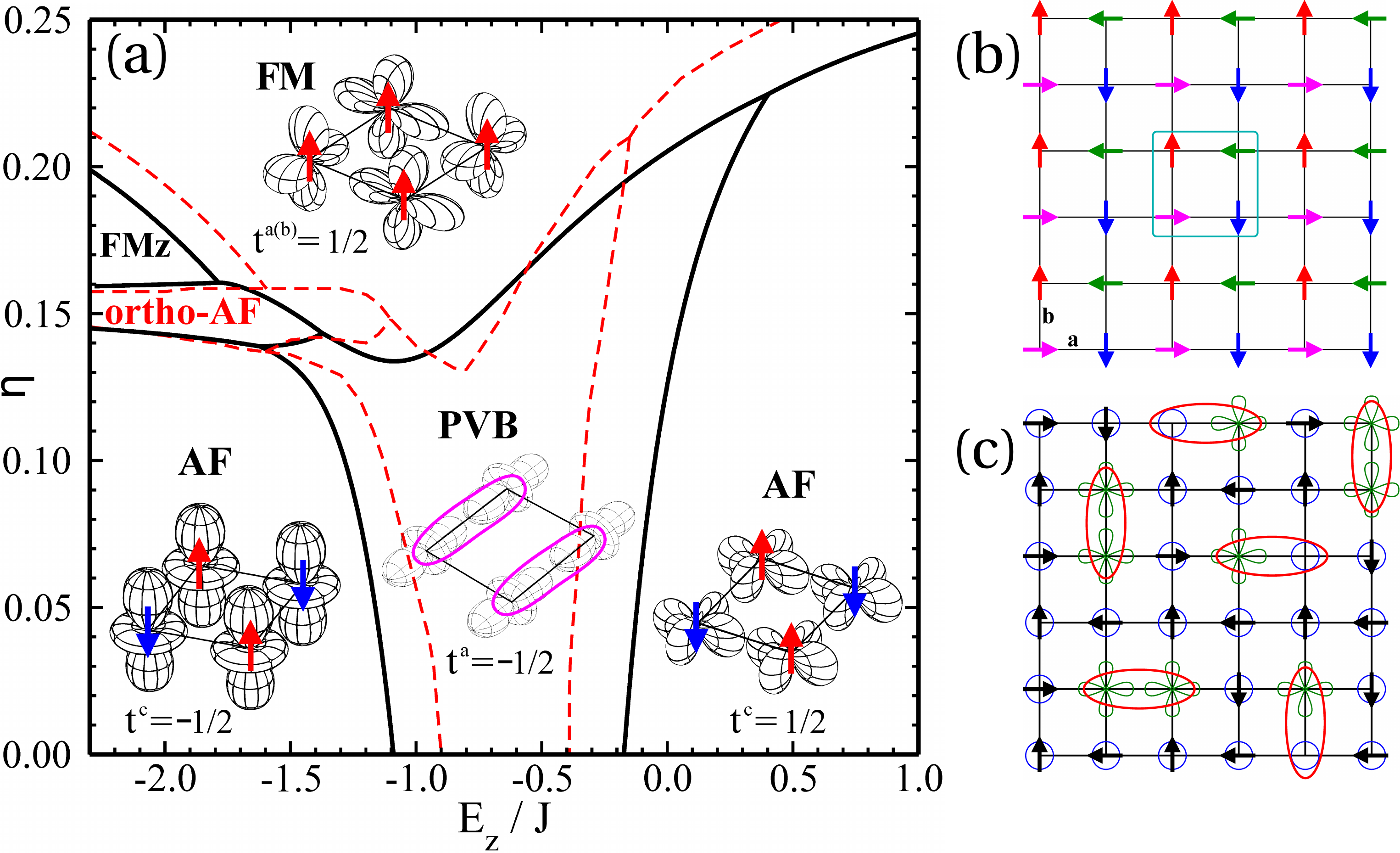}
\enc
\caption{
Spin-orbital phase diagram and entanglement in the 2D KK model:\hfill \break
(a) phase diagram in the plaquette mean field (solid lines) and ERA
(dashed lines) variational approximation, with
insets showing representative spin and orbital configurations on a
$2\times2$ plaquette --- ${\bar z}$-like
$\left(t^c\!=\!-\lla\tau^c_i\rra=\frac12\right)$ and $z$-like
$\left(t^{a,c}\!=\!-\lla\tau^{c(a)}_i\rra=-\frac12\right)$ orbitals
are accompanied either by AF long range order (arrows) or by spin
singlets on bonds in the PVB phase; \hfill\break
(b) view of an exotic four-sublattice ortho-AF phase near the
onset of FM (or FM$z$) phase; \hfill\break
(c)~artist's view of the ortho-AF phase ---
spin singlets (ovals) are entangled with either one or two orbital
excitations $|z\rangle\rightarrow|\bar{z}\rangle$ (clovers).
Images are reproduced from Ref. \cite{Brz12}.
}
\label{fig:KK_2D}
\index{quantum critical point}
\end{figure}

To capture the corrections due to quantum fluctuations, one may
construct a plaquette mean field approximation or entanglement
renormalization \textit{ansatz} (ERA) \cite{Brz12}. One finds important
corrections to a mean field phase diagram near the quantum critical
point $Q_{2{\rm D}}$, and a plaquette valence bond (PVB) state is
stable in between the above three phases with long range order, with
spin singlets on the bonds $\parallel\!a$ $\left(\parallel\!b\right)$,
stabilized by the directional orbitals $|\zeta_a\rangle$
$\left(|\zeta_b\rangle\right)$. A~novel ortho-AF phase appears as well
when the magnetic interactions change from AF to FM ones due to
increasing Hund's exchange $\eta$, and for $E_z/J<-1.5$,
see Fig. \ref{fig:KK_2D}(a). Since the nearest neighbor magnetic
interactions are very weak, exotic four-sublattice ortho-AF spin order
is stabilized by second and third nearest neighbor interactions, shown
in Fig. \ref{fig:KK_2D}(b).
Such further neighbor interactions follow from spin-orbital excitations
shown in Fig. \ref{fig:KK_2D}(c). Note that both approximate methods
employed in Ref. \cite{Brz12} (plaquette mean field approximation and
ERA) give very similar range of stability of ortho-AF phase.

\begin{figure}[t!]\bec (a)
\includegraphics[width=7.3cm,clip=true]{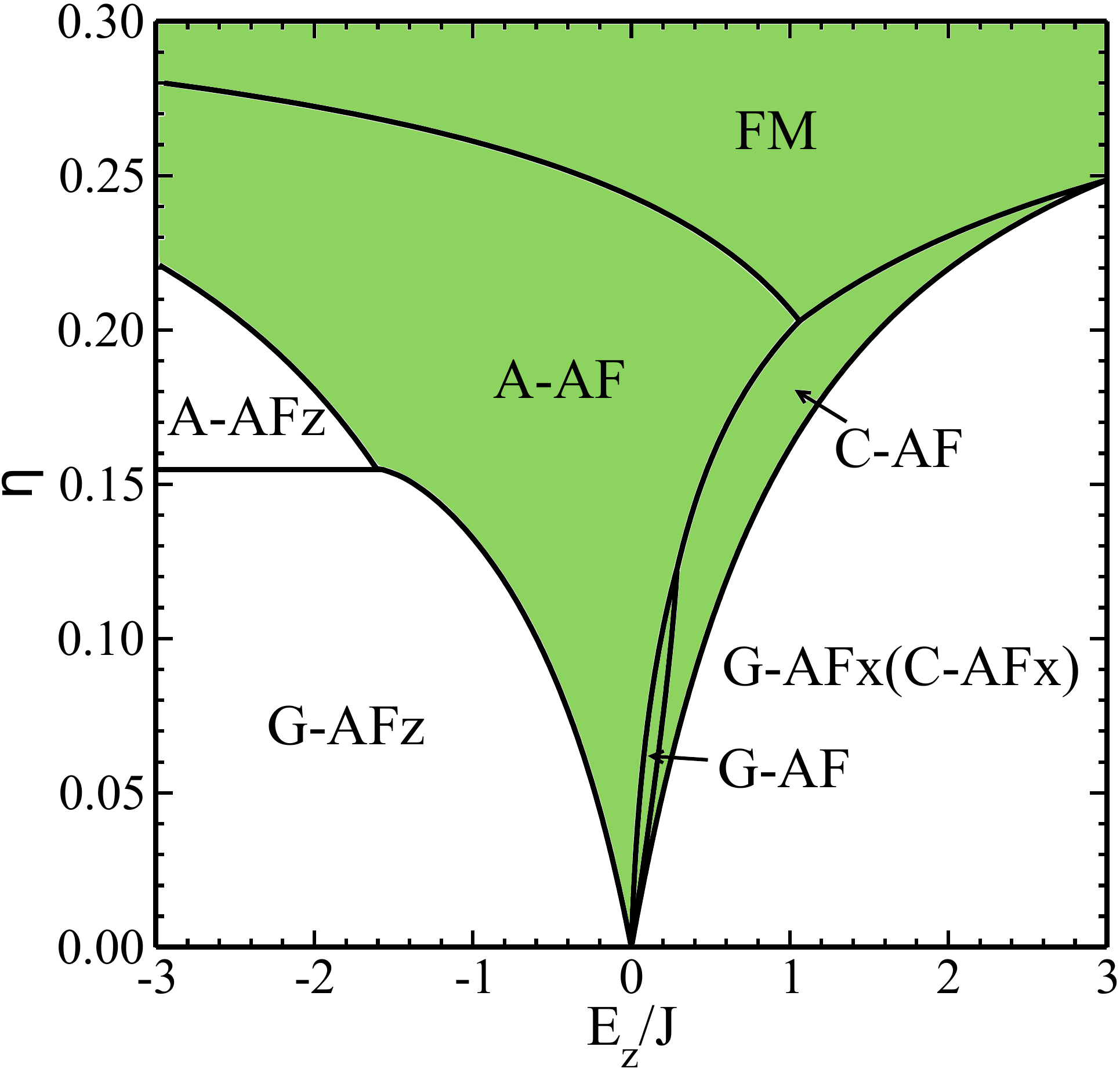}
\hskip .3cm
\includegraphics[width=7.4cm,clip=true]{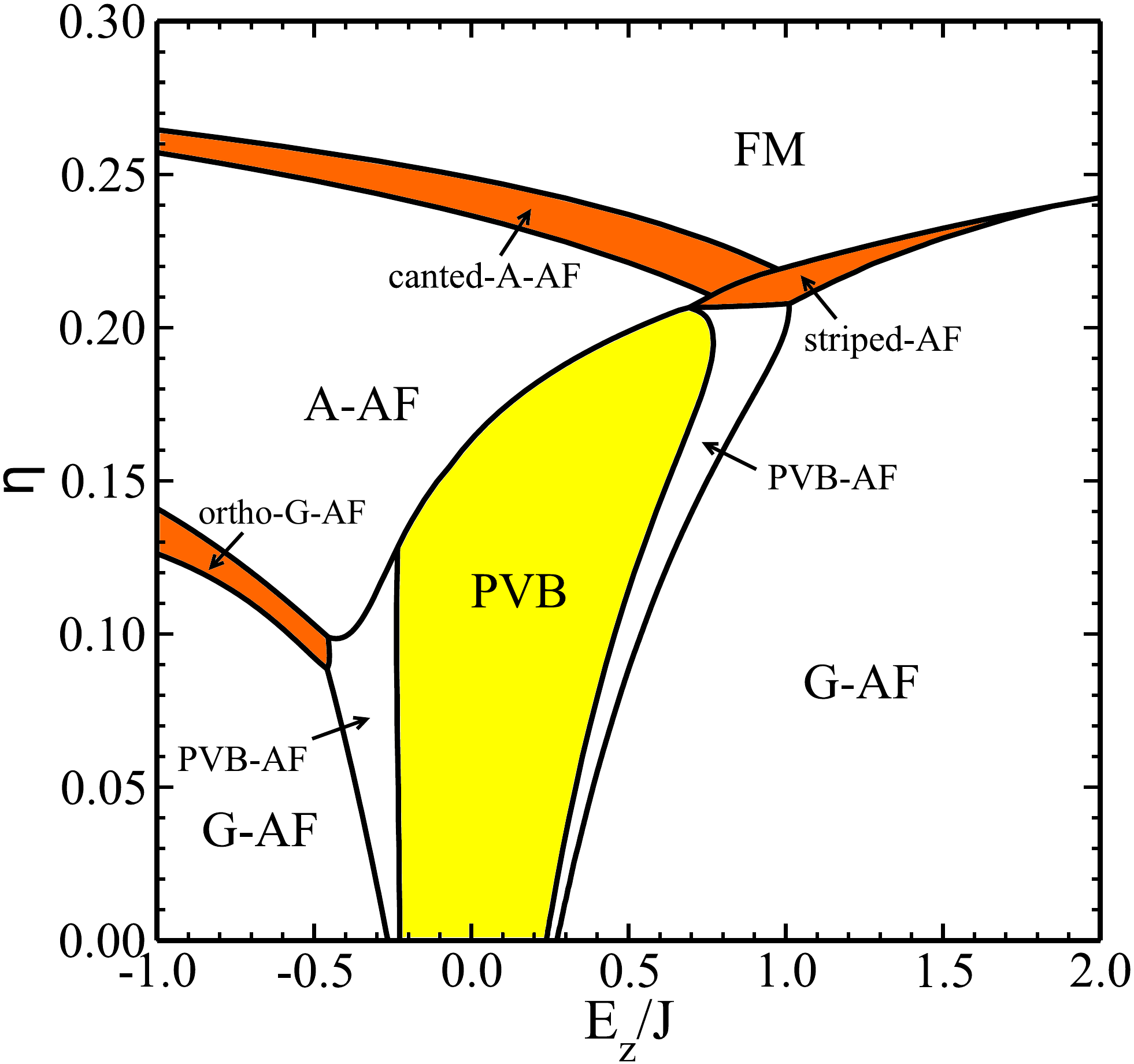}
\hskip -.4cm (b)\enc
\caption{Phase diagram of the 3D KK model obtained in two mean field
methods: (a) the single-site mean field, and
(b) the cluster mean field.
Shaded (green) area indicates phases with AO order while
the remaining magnetic phases are accompanied by FO order with fully
polarized orbitals, either ${\bar{z}}$ ($x$) (for $E_z>0$) or $z$
(for $E_z<0$). In this approach plaquette valence-bond (PVB) phase
with alternating spin singlets in the $ab$ planes (yellow) separates
the phases with magnetic long range order, see Fig. \ref{fig:so}.
Phases with exotic magnetic order are shown in orange. Note different
ranges of $E_z/J$ shown.
Images are reproduced from Ref. \cite{Brz13}.}
\label{fig:KK}
\index{quantum critical point}
\end{figure}

In the 3D KK model the exchange interaction in the $ab$ planes
(\ref{jab9}) and along the $c$ axis (\ref{jc9}) are exactly balanced at
the orbital degeneracy ($E_z=0$) and the quantum critical point where
several classical phases meet in mean field approximation is
$Q_{3{\rm D}}=(0,0)$, see Fig. \ref{fig:KK}(a). While finite $E_z$
favors one or the other $G$-AF phase, finite Hund's exchange $\eta$
favors AO order stabilizing $A$-AF spin order, see Fig. \ref{fig:so}(i).
This phase is indeed found in
KCuF$_3$ at low temperature $T<T_{\rm N}$ and is also obtained from the
electronic structure calculations \cite{Pav08}. We remark that for
unrealistically large $\eta>0.2$, spin order changes to FM.

Large qualitative changes in the phase diagram are found when spin
correlations on bonds are treated in cluster mean field approximation
(using plaquettes or linear clusters \cite{Brz13}),
see Fig.~\ref{fig:KK}(b). Phases with long range spin order ($G$-AF,
$A$-AF, and FM) are again separated by exotic types of magnetic order
which arise by a similar mechanism to that described above for an $ab$
monolayer, i.e., nearest neighbor exchange changes
sign along one cubic direction. Near the QCP $Q_{3{\rm D}}$ one finds
again PVB phase, as in the 2D KK model. In addition to the phase
diagram of Fig. \ref{fig:KK_2D}(a), the transitions between $G$-AF and
PVB phases are continuous and mixed PVB-AF phases arise.

\subsection{Spin-orbital superexchange model for LaMnO$_3$}
\label{lamno}

Electronic structure calculations give $A$-AF spin order, in agreement
with experiment. It follows from the spin-orbital superexchange for
spins $S=2$ in LaMnO$_3$, ${\cal H}_e$, due to the excitations
involving $e_g$ electrons. The energies of the five possible excited
states \cite{Griff} shown in Fig.~\ref{fig:exci}(a) are:
($i$) the HS (${\cal S}=\frac52$) $^6\!A_1$ state, and ($ii$) the LS
(${\cal S}=\frac32$) states: $^4\!A_1$,~$^4E$ ($^4E_{\epsilon}$,
$^4E_{\theta}$), and $^4\!A_2$, will be parameterized again by the
intraorbital Coulomb element $U$ and by Hund's exchange $J_H^e$ between
a pair of $e_g$ electrons in a Mn$^{2+}$ ($d^5$) ion, defined in
Eq.~(\ref{JHe}). The Racah parameters $B=0.107$ eV and $C=0.477$ eV
justify an approximate relation $C\simeq 4B$, and we find the LS
excitation spectrum:
$\varepsilon(^4\!A_1)=U+\frac34 J_H$,
$\varepsilon(^4E)=U+\frac54 J_H$ (twice), and
$\varepsilon(^4\!A_2)=U+\frac{13}{4}J_H$.

\begin{figure}[t!]
\bec (A)
\includegraphics[width=7.1cm,clip=true]{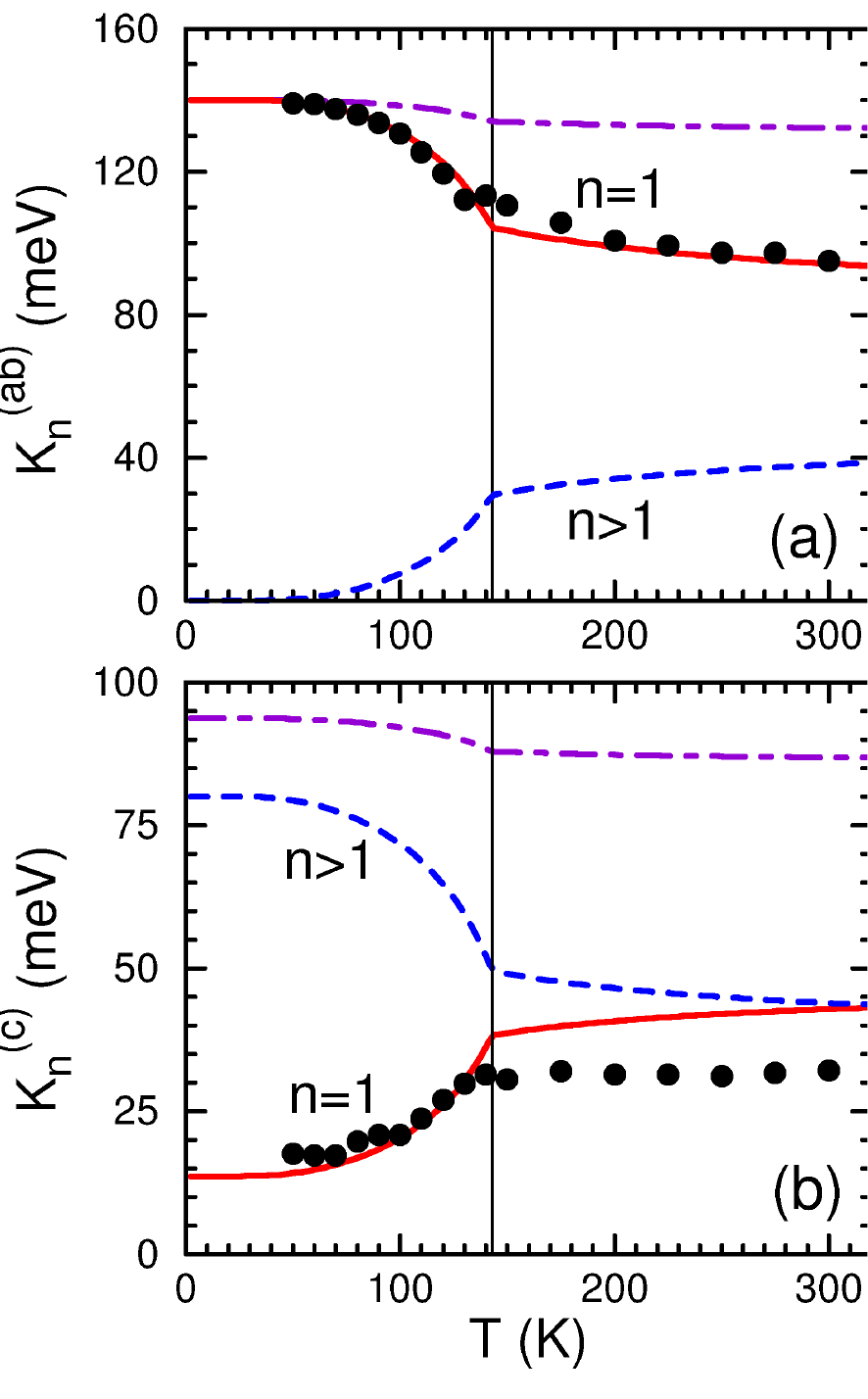} \hskip .5cm
\includegraphics[width=7.1cm,clip=true]{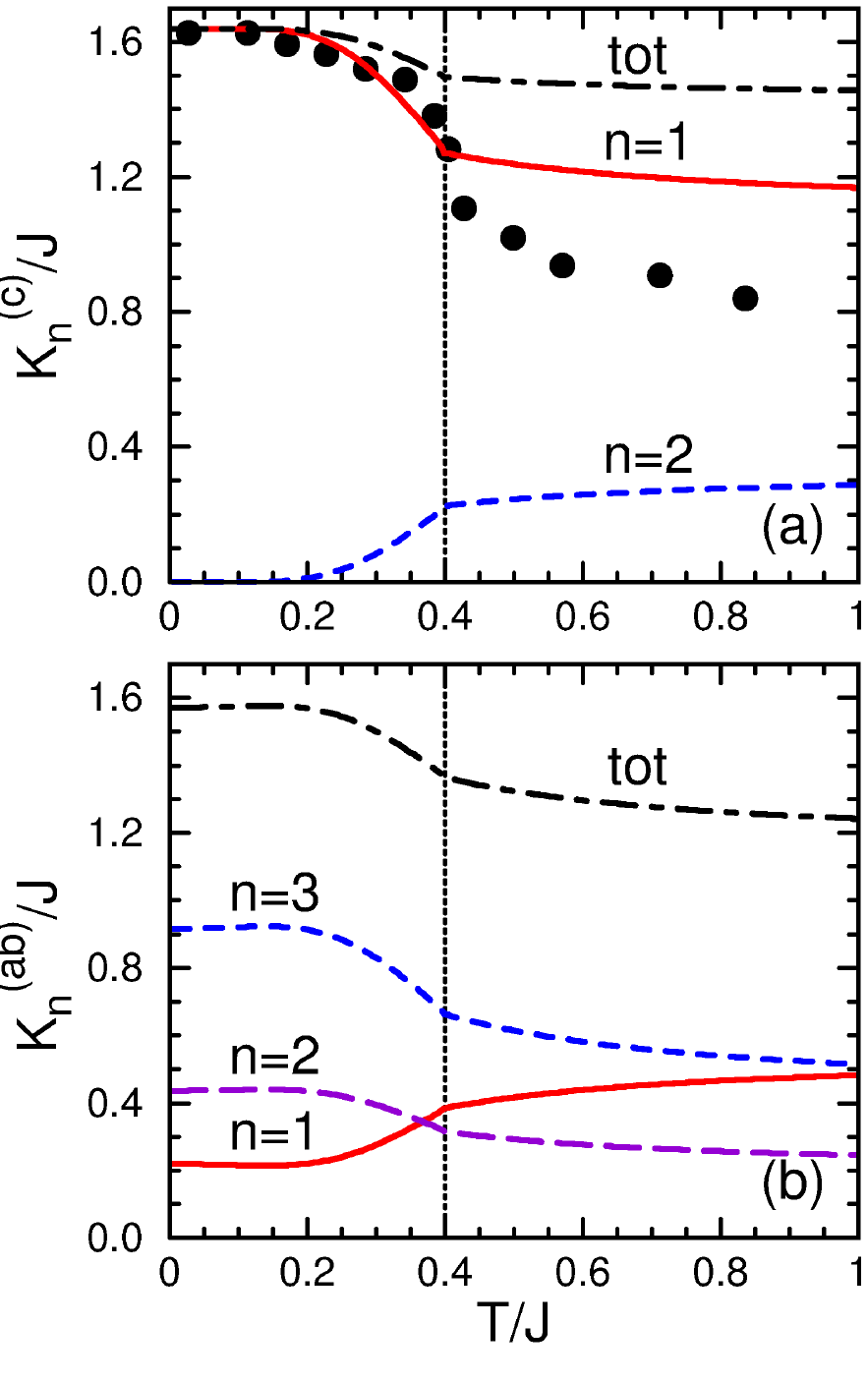} \hskip -.1cm (B)
\enc
\caption{
Kinetic energies per bond $K_n^{(\gamma)}$ (\protect\ref{hefa}) for
increasing temperature $T$ obtained from the respective spin-orbital
models for FM (top) and AF (bottom) bonds along the axis $\gamma$:
\hfill \break
(A) LaMnO$_3$ (with $J=150$ meV, $\eta\simeq 0.18$ \cite{Ole05},
end experimental points \cite{Kov10}); \hfill \break
(B) LaVO$_3$ with $\eta\!=\! 0.13$ \cite{Kha04} and
experimental points \cite{Miy02}.\hfill \break
The kinetic energies in HS states ($n=1$, red lines)
and compared with the experiment (filled circles).
Vertical dotted lines indicate the value of $T_N$.
Images are reproduced from Ref. \cite{Ole05}.
}
\label{fig:optic}
\end{figure}

Using the spin algebra (Clebsch-Gordan coefficients) and considering
again two possible $e_g$ orbital configurations, see Eqs.~(\ref{porbit})
and (\ref{qorbit}), and charge excitations by $t_{2g}$ electrons, one
finds a compact expression \cite{Fei99},
\index{spin-orbital superexchange!for LaMnO$_3$ }
\begin{eqnarray}
\label{egterm}
{\cal H}_e\!\!&=&\!\!\frac{1}{16}\sum_{\gamma}\!
\sum_{\langle ij\rangle\parallel\gamma}\left\{
 - \frac{8}{5} \frac{t^2}{\varepsilon(^6A_1)}
   \left(\vec{S}_i\cdot\vec{S}_j+6\right)
   {\cal P}_{\langle ij\rangle}^{(\gamma)}
+ \left[\frac{t^2}{\varepsilon(^4E)}
   + \frac{3}{5}\frac{t^2}{\varepsilon(^4A_1)} \right]
   \left(\vec{S}_i\cdot\vec{S}_j-4\right)
   {\cal P}_{\langle ij\rangle}^{(\gamma)}\right.       \nonumber \\
& &\left.\hskip 2.2cm +\left[ \frac{t^2}{\varepsilon(^4E)}
   + \frac{t^2}{\varepsilon(^4A_2)} \right]
   \left(\vec{S}_i\cdot\vec{S}_j-4\right)
   {\cal Q}_{\langle ij\rangle}^{(\gamma)}\right\}+E_z\sum_i\tau_i^c.\\
\label{t2gterm}
{\cal H}_t\!\!&=&\frac{1}{8}J\beta r_t
\Big(\vec{S}_i\!\cdot\!\vec{S}_j-4\Big).
\end{eqnarray}
Here $\beta=(t_{\pi}/t)^2$ follows from the difference between the
effective $d-d$ hopping elements along the $\sigma$ and $\pi$ bonds,
i.e., $\beta\simeq\frac19$, while the coefficient $r_t$ stands for a
superposition of all $t_{2g}$ excitations involved in the $t_{2g}$
superexchange \cite{Ole05}. Note that spin-projection operators for
high (low) total spin ${\cal S}=2$ (${\cal S}=1$) cannot be used, but
again the HS term stands for a FM contribution which dominates over the
other LS terms when
$\lla{\cal P}_{\langle ij\rangle}^{(\gamma)}\rra\simeq 1$.
Charge excitations by $t_{2g}$ electrons give double occupancies in
active $t_{2g}$ orbitals, so ${\cal H}_t$ is AF but this term is small
--- as a result FM interactions may dominate but again only along two
spatial directions. Indeed, this happens for the realistic parameters
of LaMnO$_3$ for the $ab$ planes where spin order is FM and coexists
with AO order, while along the $c$ axis spin order is AF accompanied by
FO order, i.e., spin-orbital order is $A$-AF/$C$-AF. Indeed, this type
of order is found both from the theory for realistic parameters and
from the electronic structure calculations \cite{Pav10}. One concludes
that Jahn-Teller orbital interactions are responsible for the enhanced
value of the orbital transition temperature \cite{Sna16}.

The optical spectral weight due to HS states in LaMnO$_3$ may be easily
derived from the present model (\ref{egterm}), following the general
theory, see Eq. (\ref{hefa}).
\index{optical spectral weight}
One finds a very satisfactory agreement between the present theory and
the experimental results of \cite{Kov10}, as shown in Fig.
\ref{fig:optic}(A). We emphasize, that no fit is made here, i.e., the
kinetic energies (\ref{hefa}) are calculated using the same parameters
as those used for the magnetic exchange constants \cite{Ole05}.
Therefore, such a good agreement with experiment suggests that indeed
the spin-orbital superexchange may be disentangled, as also verified
later \cite{Sna16}.

\begin{figure}[t!]\bec (a) \hskip -.1cm
\includegraphics[width=7.2cm,clip=true]{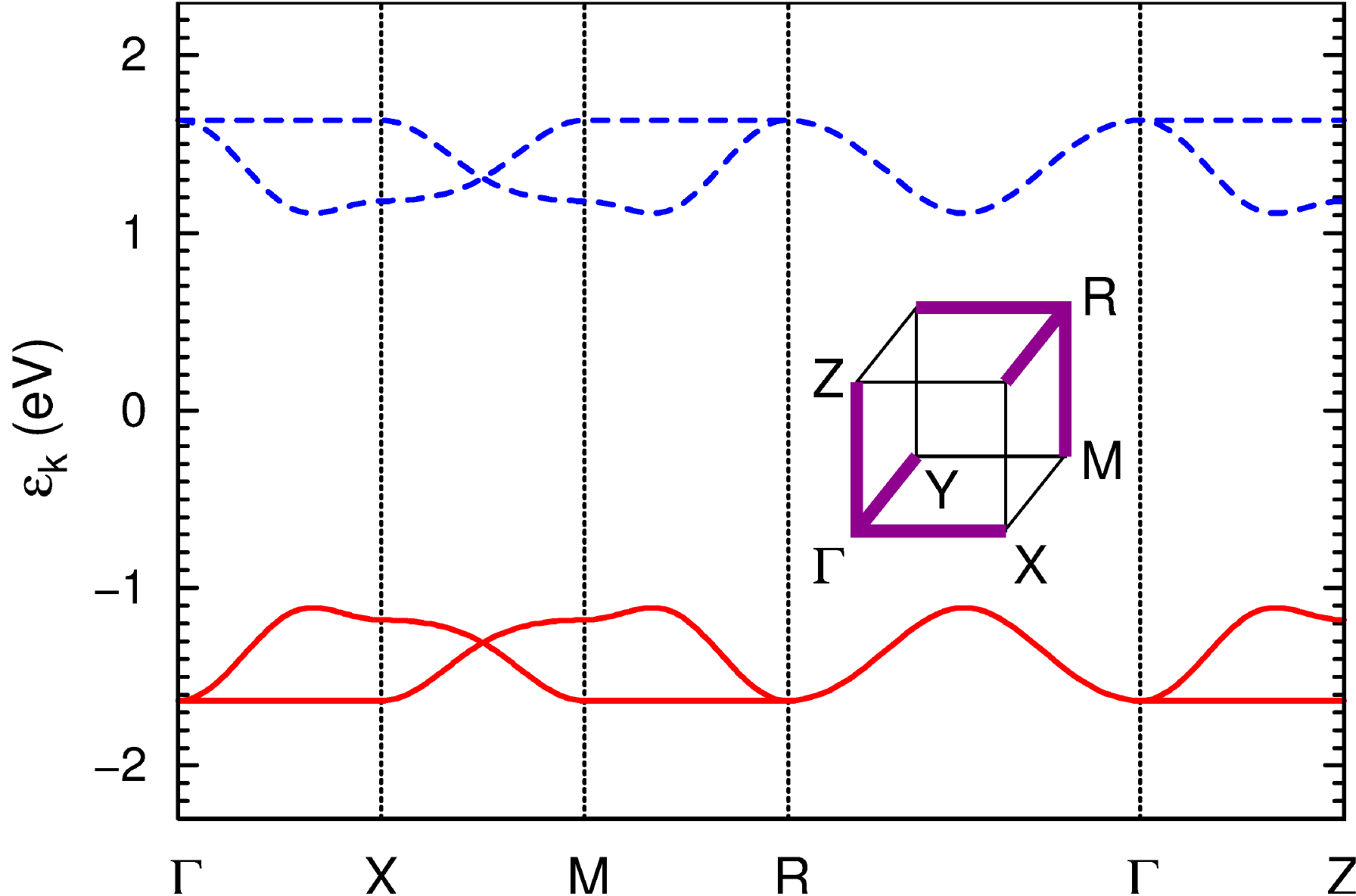}
\hskip .3cm
\includegraphics[width=7.2cm,clip=true]{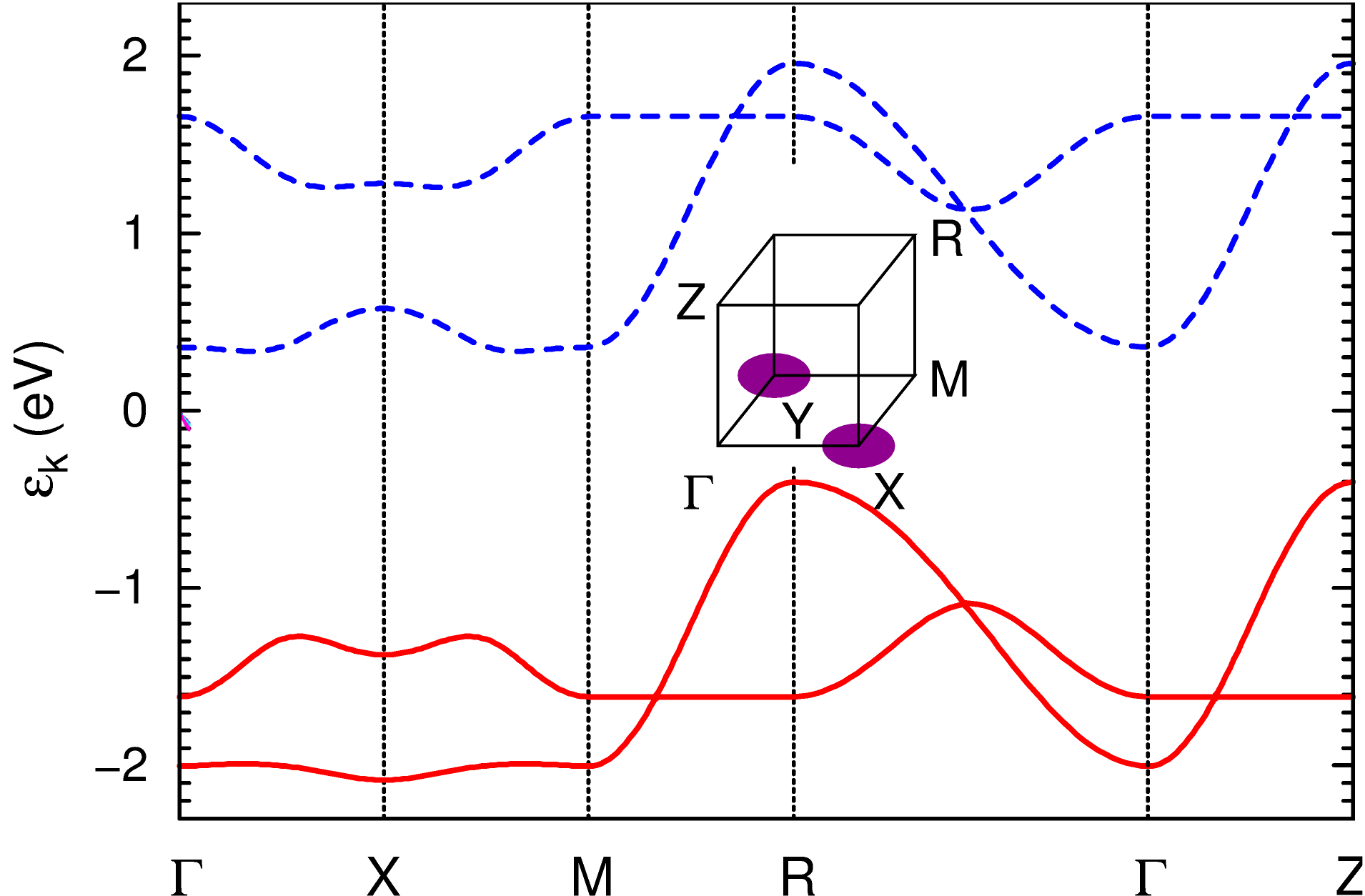}\hskip .2cm (b)
\enc
\caption{
Band structure along
the high symmetry directions in:
(a) $G$-AF phase at $x=0$ and
(b) $C$-AF phase at $x=0.05$.
Spin majority (minority) bands are shown by solid (dashed) lines.
Parameters: $t=0.4$ eV, $J_H=0.74$ eV, $g=3$ eV.
Insets shows the Fermi surfaces at low doping. The special points:
$\Gamma=(0,0,0)$, $X=(\pi,0,0)$, $M=(\pi,\pi,0)$, $R=(\pi,\pi,\pi)$,
$Z=(0,0,\pi)$.
Images are reproduced from Ref. \cite{Ole11}.}
\label{fig:ema}
\end{figure}

To give an example of a phase transition triggered by $e_g$ electron
doping of Sr$_{1-x}$La$_x$MnO$_3$ we show the results obtained with
double exchange model for degenerate $e_g$ electrons
extended by the coupling to the lattice \cite{Ole11},
\beq
{\cal H}\!=-\!\sum_{ij,\alpha\beta,\sigma} t_{\alpha\beta}^{ij}
a_{i\alpha\sigma}^{\dagger} a_{j\beta\sigma}^{}
-2J_H\sum_i {\vec S}_i\cdot{\vec s}_i
+\!J\sum_{\langle ij\rangle}{\vec S}_i\cdot{\vec S}_j
-\!gu\sum_i(n_{iz}-n_{i\bar{z}})+\!\frac12 NK u^2.
\label{de}
\eeq
It includes the hopping of $e_g$ electrons between orbitals
$\alpha=z,\bar{z}$ as in Eq. (\ref{Hxz}). The tetragonal distortion $u$
is finite only in the $C$-AF phase. Here we define it as proportional
to a difference between two lattice constants $a$ and $c$ along the
respective axis, $u\equiv 2(c-a)/(c+a)$, and $N$ is the number of
lattice sites.
The microscopic model that explains the mechanism of the magnetic
transition in electron doped manganites from canted $G$-AF to collinear
$C$-AF phase at low doping $x\simeq 0.04$. The double exchange
supported by the cooperative Jahn-Teller effect leads then to
dimensional reduction from an isotropic 3D $G$-AF phase to a quasi-1D
order of partly occupied $3z^2-r^2$ orbitals in the $C$-AF phase
\cite{Ole11}. We emphasize that this theory prediction relies on the
shape of the Fermi surface which is radically different in the $G$-AF
and $C$-AF phase. Due to the Fermi surface topology, spin canting is
suppressed in the $C$-AF phase, in agreement with the experiment.

\section{Superexchange for active $t_{2g}$ orbitals}
\label{sext}

\subsection{Spin-orbital superexchange model for LaTiO$_3$}
\label{latio}

LaTiO$_3$ would be electron-hole symmetric compound to KCuF$_3$, if not
the orbital degree of freedom which $t_{2g}$ here. This changes the
nature of orbital operators from the projections for each bond to
scalar products of pseudospin $T=\frac12$ operators. The superexchange
spin-orbital model (\ref{som}) in the perovskite titanates couples
$S=\frac12$ spins and $T=\frac12$ pseudospins arising from the $t_{2g}$
orbital degrees of freedom at nearest neighbor Ti$^{3+}$ ions, e.g. in
LaTiO$_3$ or YTiO$_3$ \cite{Kha00}. Due to large intraorbital Coulomb
element $U$ electrons localize
and the densities satisfy the local constraint at each site $i$,
\begin{equation}
n_{ia}+n_{ib}+n_{ic}=1.
\label{cond1}
\end{equation}
The charge excitations lead to one of four different excited states
\cite{Griff}, shown in Fig. \ref{fig:exci}(b): \hfill\\
($i$) the high-spin $^3T_1$ state at energy $U-3J_H$, and \hfill\\
($ii$) three low-spin states --- degenerate $^1T_2$ and $^1E$ states at
energy $(U-J_H)$, and \hfill\\
($iii$) an $^1\!A_1$ state at energy $(U+2J_H)$. \hfill\\
As before, the excitation energies are parameterized by $\eta$, defined
by Eq. (\ref{eta}), and we introduce the coefficients
\begin{equation}
r_1=\frac{1}{1-3\eta}, \hskip .7cm
r_2=\frac{1}{1- \eta}, \hskip .7cm
r_3=\frac{1}{1+2\eta}.
\label{rd1}
\end{equation}

One finds the following compact expressions for the terms contributing
to superexchange ${\cal H}_J(d^1)$ Eq. (\ref{HJ}) \cite{Kha00}:
\index{spin-orbital superexchange!for LaTiO$_3$}
\begin{eqnarray}
\label{som11}
H_1^{(\gamma)}\!&=&\!\frac{1}{2}Jr_1
    \left(\vec{S}_i\!\cdot\!\vec{S}_j+\frac{3}{4}\right)
    \left(A_{ij}^{(\gamma)}\!-\frac{1}{2}n_{ij}^{(\gamma)}\right),  \\
\label{som12}
H_2^{(\gamma)}\!&=&\! \frac{1}{2}Jr_2
    \left(\vec{S}_i\!\cdot\!\vec{S}_j-\frac{1}{4}\right)
    \left(A_{ij}^{(\gamma)}\!-\frac{2}{3}B_{ij}^{(\gamma)}\!
    +\frac{1}{2}n_{ij}^{(\gamma)}\right),                            \\
\label{som13}
H_3^{(\gamma)}\!&=&\! \frac{1}{3}Jr_3
    \left(\vec{S}_i\!\cdot\!\vec{S}_j-\frac{1}{4}\right)B_{ij}^{(\gamma)},
\end{eqnarray}
where
\beq
A_{ij}^{(\gamma)}\!=
2\left(\vec{\tau}_i\cdot\vec{\tau}_j+\frac{1}{4}n_in_j\right)^{(\gamma)}\!,
                                                 \hskip .3cm
B_{ij}^{(\gamma)}\!=
2\left(\vec{\tau}_i\otimes\vec{\tau}_j+\frac{1}{4}n_in_j\right)^{(\gamma)}\!,
                                                 \hskip .3cm
n_{ij}^{(\gamma)}\!=n_{i}^{(\gamma)}+n_{j}^{(\gamma)}.
\eeq
As in Sec. \ref{sec:KK}, the orbital (pseudospin) operators
$\left\{A_{ij}^{(\gamma)},B_{ij}^{(\gamma)},n_{ij}^{(\gamma)}\right\}$
depend on the direction of the $\langle ij\rangle\parallel\gamma$ bond.
Their form follows from two active $t_{2g}$ orbitals (flavors) along
the cubic axis $\gamma$, e.g. for $\gamma=c$ the active orbitals are
$a$ and $b$, and they give two components of the pseudospin $T=\frac12$
operator $\vec{\tau}_i$. The operators
$\left\{A_{ij}^{(\gamma)},B_{ij}^{(\gamma)}\right\}$ describe the
interactions between these two active orbitals, which include the
quantum fluctuations, and take either the form of a scalar product
$\vec{\tau}_i\cdot\vec{\tau}_j$ in $A_{ij}^{(\gamma)}$, or lead to
a similar expression,
\begin{equation}
\label{tauxtau}
\vec{\tau}_i\otimes\vec{\tau}_j=
\tau_i^x\tau_j^x-\tau_i^y\tau_j^y+\tau_i^z\tau_i^z,
\end{equation}
in $B_{ij}^{(\gamma)}$. These latter terms
enhance orbital fluctuations by double excitations due to the
$\tau_i^+\tau_j^+$ and $\tau_i^-\tau_j^-$ terms. The interactions along
the axis $\gamma$ are tuned by the number of electrons occupying active
orbitals, $n_i^{(\gamma)}=1-n_{i\gamma}$, which is fixed by the number
of electrons in the inactive orbital $n_{i\gamma}$ by the constraint
(\ref{cond1}). The cubic titanates are known to have particularly
pronounced quantum spin-orbital fluctuations \cite{Kha05}, and their
proper treatment requires a rather sophisticated approach. Therefore,
in contrast to AF long range order found in $e_g$-orbital systems,
spin-orbital disordered state may occur in titanium perovskites, as
suggested for LaTiO$_3$~\cite{Kha00}.

\subsection{Spin-orbital superexchange model for LaVO$_3$}
\label{lavo}

As the last cubic system we present the spin-orbital model for V$^{3+}$
ions in $d^2$ configurations in the vanadium perovskite $R$VO$_3$
($R$=La,$\dots$,Lu). Due to Hund's exchange one has
$S=1$ spins and three ($n=1,2,3$) charge excitations $\varepsilon_n$
arising from the transitions to [see Fig. \ref{fig:exci}(b)]:\hfill \\
  ($i$) a high-spin state $^4\!A_2$ at energy $(U-3J_H)$,\hfill \\
 ($ii$) two degenerate low-spin states $^2T_1$ and $^2E$ at $U$, and\hfill \\
($iii$) $^2T_2$ low-spin state at $(U+2J_H)$ \cite{Kha01}.\hfill \\
Using $\eta$ (\ref{eta}) we parameterize this multiplet structure by
\beq
r_1=\frac{1}{1-3\eta}, \hskip 1cm r_3=\frac{1}{1+2\eta}.
\eeq
The cubic symmetry is broken and the crystal field induces orbital
splitting in $R$VO$_3$, hence
$\langle n_{ic}\rangle=1$ and the orbital degrees of freedom are given
by the doublet $\{a,b\}$ which defines the pseudospin operators
${\vec\tau}_i$ at site $i$. One derives a HS contribution
$H_1^{(c)}(ij)$ for a bond ${\langle ij\rangle}$ along the $c$ axis,
and $H_1^{(ab)}(ij)$ for a bond in the $ab$ plane:
\index{spin-orbital superexchange!for LaVO$_3$}
\bea
\label{H1c}
H_1^{(c)}(ij)&=&-\frac{1}{3}Jr_1\left(\vec S_i\!\cdot\!\vec S_j+2\right)
\left(\textstyle{\frac14}-\vec\tau_i\!\cdot\!\vec\tau_j\right), \\
\label{H1ab}
H_1^{(ab)}(ij)&=&-\frac{1}{6}Jr_1\left(\vec S_i\!\cdot\!\vec S_j+2\right)
\left(\textstyle{\frac14}-\tau_i^z\tau_j^z\right).
\eea
In Eq. (\ref{H1c}) pseudospin operators $\vec\tau_i$ describe
low-energy dynamics of (initially degenerate) $\{xz,yz\}$ orbital
doublet at site $i$; this dynamics is quenched in $H_1^{(ab)}$
(\ref{H1ab}).
Here $\frac{1}{3}(\vec S_i\cdot\vec S_j+2)$ is the projection operator
on the HS state for $S=1$ spins. The terms $H_n^{(c)}(ij)$ for LS
excitations ($n=2,3$) contain instead the spin operator
$(1-\vec S_i\cdot\vec S_j)$ (which guarantees that these terms cannot
contribute for fully polarized spins $\langle\vec S_i\cdot\vec S_j\rangle=1$):
\begin{eqnarray}
\label{H23c}
H_2^{(c)}(ij)&=&-\frac{1}{12}J \left(1-\vec S_i\!\cdot\!\vec S_j\right)
\left(\textstyle{\frac{7}{4}}-\tau_i^z \tau_j^z - \tau_i^x \tau_j^x
+5\tau_i^y \tau_j^y\right),                              \nonumber \\
H_3^{(c)}(ij)&=&-\frac{1}{4}Jr \left(1-\vec S_i\!\cdot\!\vec S_j\right)
\left(\textstyle{\frac{1}{4}}+\tau_i^z \tau_j^z+\tau_i^x \tau_j^x
-\tau_i^y \tau_j^y\right),
\end{eqnarray}
while again the terms $H_n^{(ab)}(ij)$ differ from $H_n^{(c)}(ij)$
only by orbital operators:
\begin{eqnarray}
\label{H23ab}
H_2^{(ab)}(ij)&=&-\frac{1}{8}J\left(1-\vec S_i\!\cdot\!\vec S_j\right)
\left(\textstyle{\frac{19}{12}}\mp \textstyle{\frac{1}{2}}\tau_i^z
\mp \textstyle{\frac{1}{2}}\tau_j^z
-\textstyle{\frac{1}{3}}\tau_i^z\tau_j^z\right),             \nonumber \\
H_3^{(ab)}(ij)&=&-\frac{1}{8}Jr\left(1-\vec S_i\!\cdot\!\vec S_j\right)
\left(\textstyle{\frac{5}{4}}\mp \textstyle{\frac{1}{2}}\tau_i^z
\mp \textstyle{\frac{1}{2}}\tau_j^z+\tau_i^z\tau_j^z\right),
\end{eqnarray}
where upper (lower) sign corresponds to bonds along the $a$($b$) axis.

First we present a mean field approximation for the spin and orbital
bond correlations which are determined self-consistently after
decoupling them from each other in ${\cal H}_J$ (\ref{som}). Spin
interactions in Eq. (\ref{Hs}) are given by two exchange constants:
\begin{eqnarray}
J_c &=&
\frac{1}{2}J\left\{\eta r_1-(r_1-\eta r_1-\eta r_3)
(\textstyle{\frac{1}{4}}+\langle\vec \tau_i\!\cdot\!\vec\tau_j\rangle)
-2\eta r_3 \langle \tau_i^y \tau_j^y \rangle\right\},     \nonumber  \\
J_{ab} &=&\frac{1}{4}J\left\{1-\eta r_1-\eta r_3+
(r_1-\eta r_1-\eta r_3)(\textstyle{\frac{1}{4}}
+\langle\tau_i^z\tau_j^z\rangle)\right\},
\label{Jspin}
\end{eqnarray}
determined by orbital correlations
$\langle\vec \tau_i\!\cdot\!\vec\tau_j\rangle$ and
$\langle\tau_i^{\alpha}\tau_j^{\alpha}\rangle$. By evaluating them one
 finds $J_c<0$ and $J_{ab}>0$ supporting $C$-AF
spin order. In the orbital sector one finds
\begin{equation}
\label{Horb}
H_{\tau}=
\sum_{\langle ij\rangle_{c}}\left[J_c^{\tau}\vec\tau_i\cdot\vec\tau_j
-J(1-s_c)\eta r_3\tau_i^y\tau_j^y\right]
+J_{ab}^{\tau}\sum_{\langle ij\rangle_{ab}}\tau_i^z\tau_j^z,
\end{equation}
with:
\begin{eqnarray}
J_c^{\tau}&=&
\frac{1}{2}J\left[(1+s_c)r_1+(1-s_c)\eta(r_1+r_3)\right],        \nonumber  \\
J_{ab}^{\tau}&=&
\frac{1}{4}J\left[(1-s_{ab})r_1+(1+s_{ab})\eta(r_1+r_3)\right],
\label{Jorb}
\end{eqnarray}
depending on spin correlations:
$s_c=\langle \vec S_i\cdot\vec S_j\rangle_c$ and
$s_{ab}=-\langle\vec S_i\cdot\vec S_j\rangle_{ab}$. In a classical
$C$-AF state ($s_c=s_{ab}=1$) this mean field procedure becomes exact,
and the orbital problem maps to Heisenberg pseudospin chains along the
$c$ axis, weakly coupled (as $\eta\ll 1$) along $a$ and $b$ bonds,
\begin{equation}
\label{Horb0}
H_{\tau}^{(0)}=Jr_1\left[\sum_{\langle ij\rangle_c}\vec\tau_i\cdot\vec\tau_j
+\frac{1}{2}\eta\left(1+\frac{r_3}{r_1}\right)\sum_{\langle ij\rangle_{ab}}
\tau_i^z\tau_j^z\right],
\end{equation}
releasing large zero-point energy. Thus, spin $C$-AF and $G$-AO order
with quasi-1D orbital quantum fluctuations support each other in
$R$VO$_3$. Orbital fluctuations play here a prominent role and
amplify the FM exchange $J_c$, making it even stronger that the AF
exchange $J_{ab}$ \cite{Kha01}.

Having the individual terms $H_n^{(\gamma)}$ of the spin-orbital model,
one may derive the spectral weights of optical spectra (\ref{hefa}).
The HS excitations have remarkable temperature dependence and the
spectral weight decreases in the vicinity of the magnetic transition
at $T_{\rm N}$, see Fig. \ref{fig:optic}(B). The observed behavior is
reproduced in the theory only when spin-orbital interactions are
treated in a cluster approach, i.e. they \textit{cannot} be disentangled,
see Sec. \ref{sec:enta}.
\index{optical spectral weight}

\begin{figure}[t!]\bec (a) \hskip -.5cm
\includegraphics[width=7.2cm,clip=true]{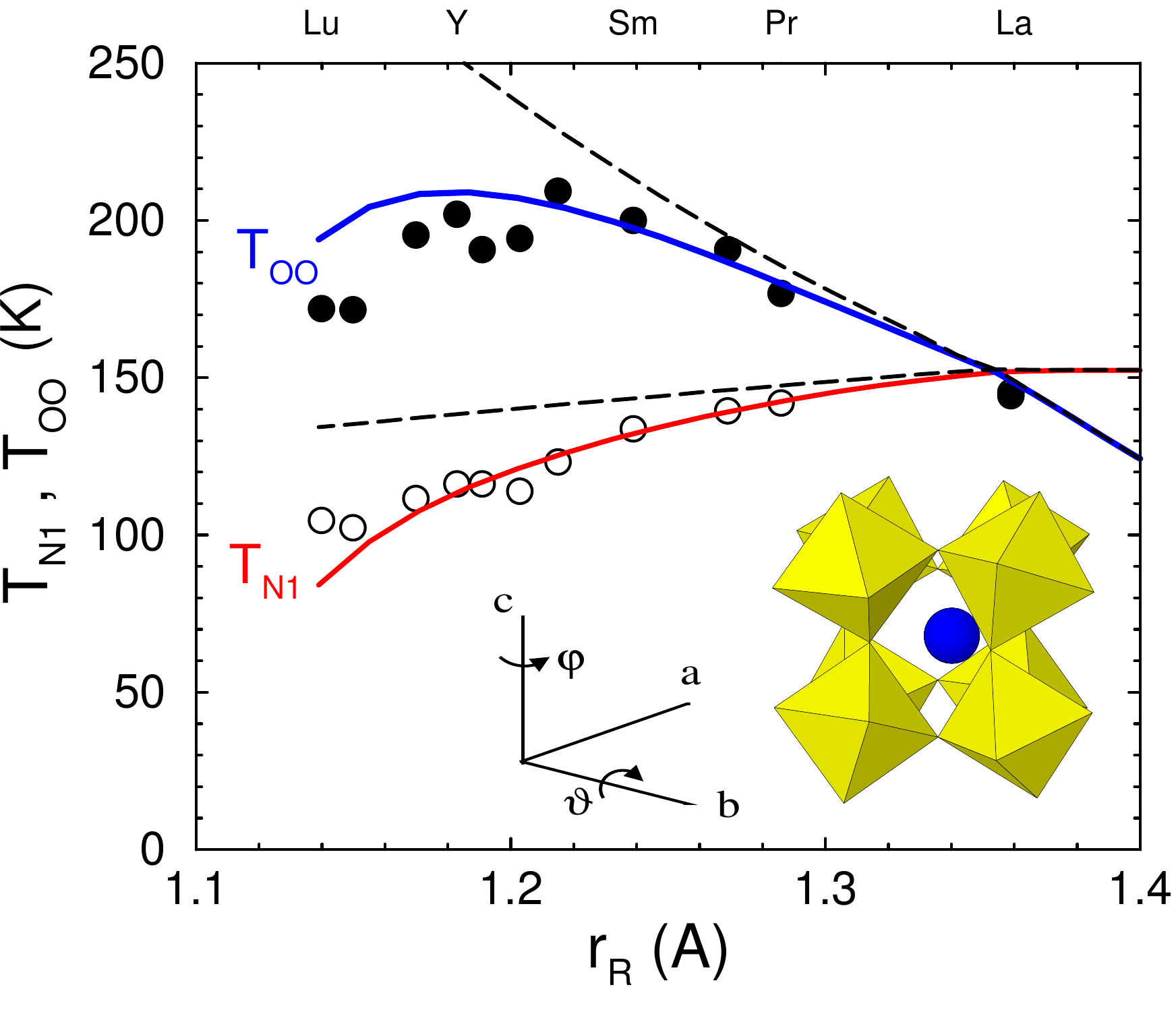}
\hskip .3cm
\includegraphics[width=8.0cm,clip=true]{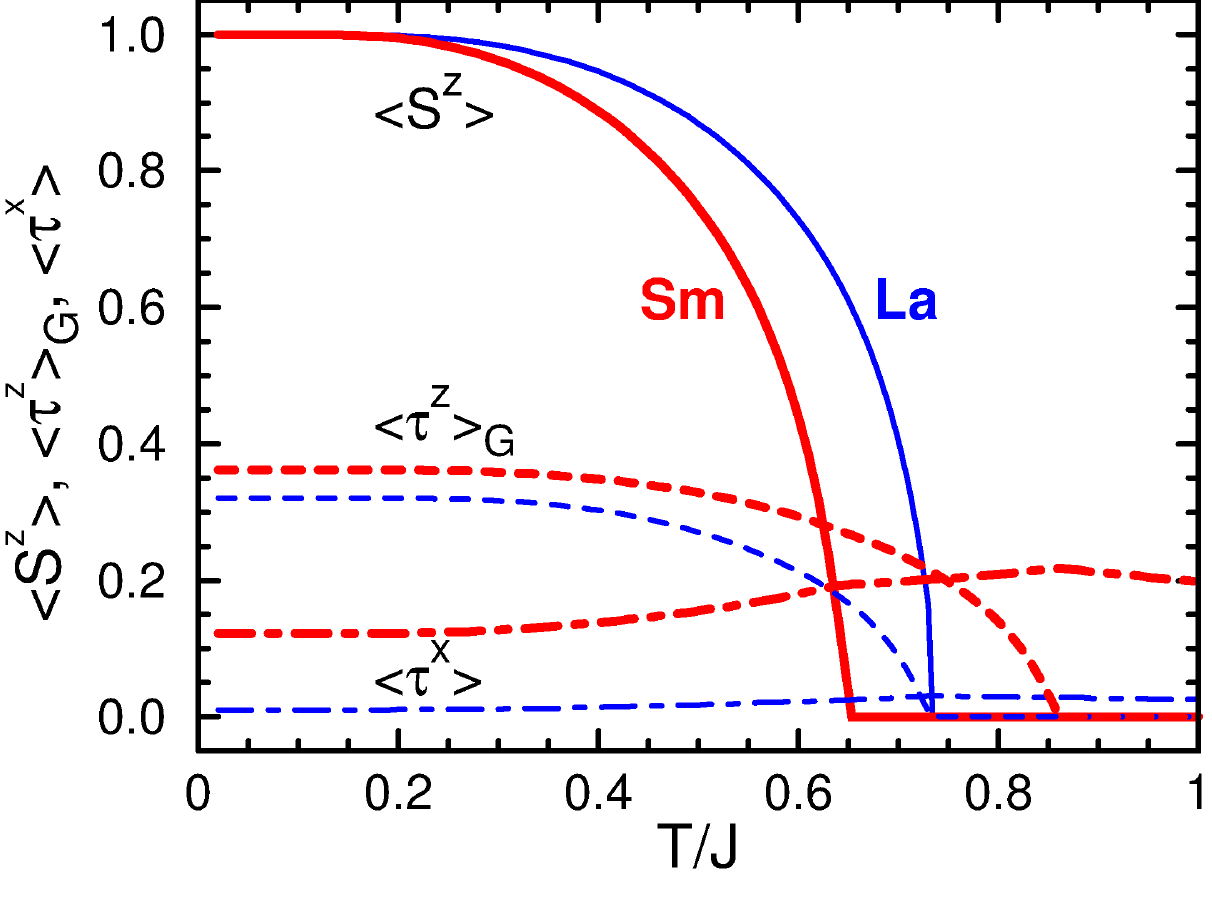} \hskip -.5cm (b)
\enc
\caption{Phase transitions in the vanadium perovskites $R$VO$_3$:
(a) phase diagram with the orbital $T_{\rm OO}$ and N\'eel $T_{N1}$
transition temperature obtained from the theory with and without
orbital-lattice coupling (solid and dashed lines)
\cite{Hor08}, and from experiment (circles) \cite{Miy03}; \hfill\break
(b) spin $\langle S_i^z\rangle$ (solid) and $G$-type orbital
$\langle\tau_i^z\rangle_G$ (dashed) order parameters, vanishing at
$T_{\rm OO}$ and $T_{N1}$, and the transverse orbital polarization
$\langle\tau_i^x\rangle$ (dashed-dotted lines) for LaVO$_3$ and
SmVO$_3$ (thin and heavy lines).
Images are reproduced from Ref. \cite{Hor08}.
}
\label{fig:phd}
\end{figure}

Unlike in LaMnO$_3$ where the spin and orbital phase transitions are
well separated, in the $R$VO$_3$ ($R$=Lu,Yb,$\dots$,La) the two
transitions are close to each other \cite{Miy03}. It is not easy to
reproduce the observed dependence of the transition temperatures
$T_{\rm OO}$ and N\'eel $T_{N1}$ on the ionic radius $r_R$ (in the
$R$VO$_3$ compounds with small $r_R$ there is also another magnetic
transition at $T_{N2}$ \cite{Fuj10} which is not discussed here).
The spin-orbital model was extended by the coupling to the lattice to
unravel a nontrivial interplay between superexchange, the
orbital-lattice coupling due to the GdFeO$_3$-like rotations of the
VO$_6$ octahedra, and orthorhombic lattice distortions \cite{Hor08}.
One finds that the lattice strain affects the onset of the magnetic
and orbital order by partial suppression of orbital fluctuations,
and the dependence of $T_{\rm OO}$ is non-monotonous in
Fig.~\ref{fig:phd}(a). Thereby the orbital polarization
$\propto\lla\tau^x\rra$ increases with decreasing ionic radius $r_R$,
and the value of $T_{N1}$ is reduced, see Fig. \ref{fig:phd}(b).
The theoretical approach demonstrates that orbital-lattice coupling
is very important and reduces both $T_{\rm OO}$ and N\'eel $T_{N1}$
for small ionic radii.

\section{Spin-orbital complementarity and entanglement}
\label{soe}

\subsection{Goodenough-Kanamori rules}
\index{Goodenough-Kanamori rules}

While rather advanced many-body treatment of the quantum physics
characteristic for spin-orbital models is required in general, we want
to present here certain simple principles which help to understand the
heart of the problem and to give simple guidelines for interpreting
experiments and finding relevant physical parameters of the
spin-orbital models of {\it undoped\/} cubic insulators. We will argue
that such an approach based upon classical orbital order is well
justified in many known cases, as quantum phenomena are often quenched
by the Jahn-Teller (JT) coupling between orbitals and the lattice
distortions, which are present below structural phase transitions and
induce orbital order both in spin-disordered and in spin-ordered or
spin-liquid phases.

\begin{figure}[t!]
\bec
\includegraphics[width=12cm,clip=true]{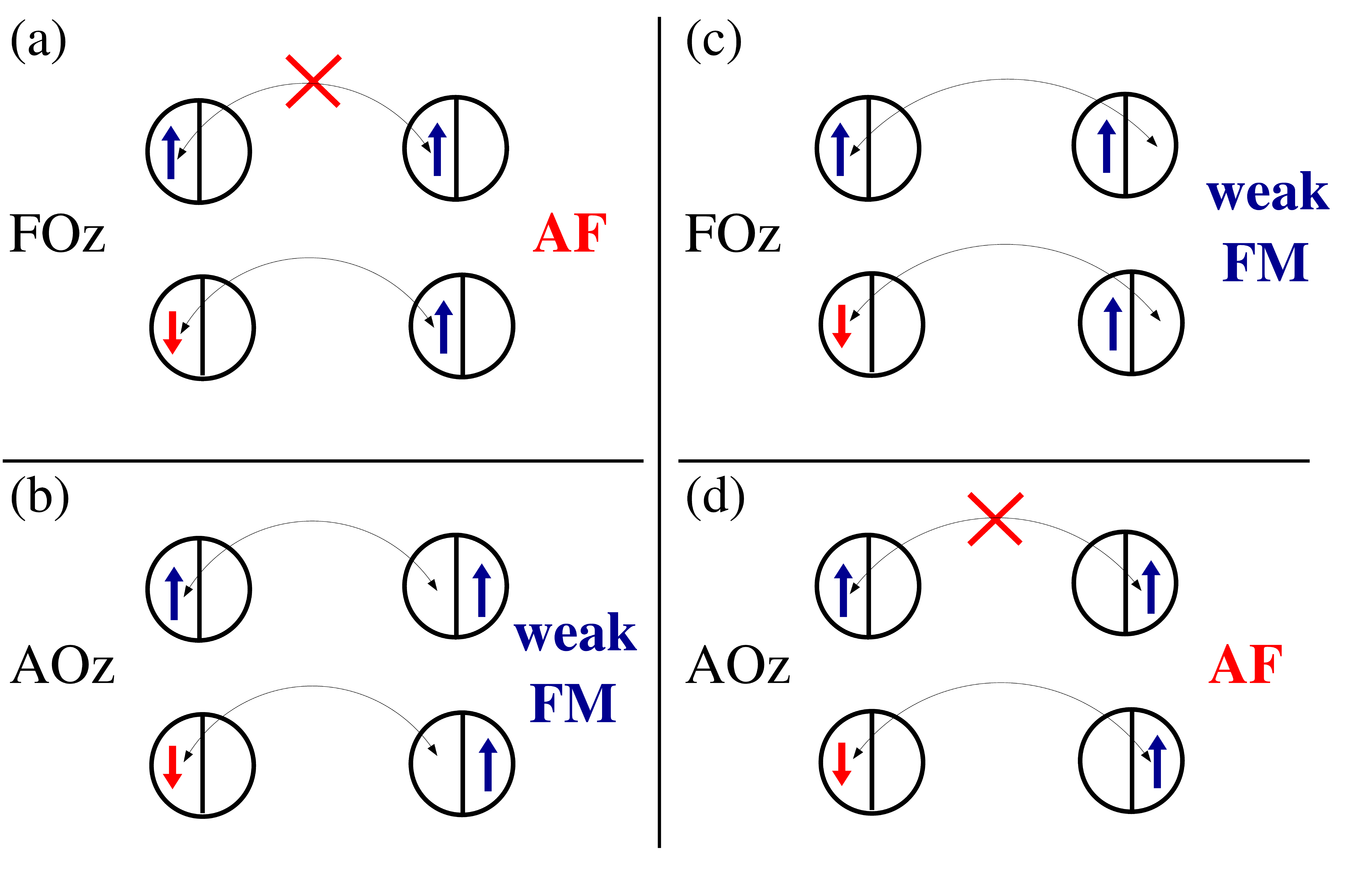}
\enc
\caption{Artist's view of the GKR \cite{Goode} for:
(a) FO$z$ and AF spin order and
(b) AO$z$ and FM spin order in a system with orbital flavor conserving
hopping as is alkali $R$O$_2$ hyperoxides ($R$=K,Rb,Cs) \cite{Hyper}.
The charge excitations generated by interorbital hopping fully violate
the GKR and support the states with the same spin-orbital order:
(c) FO$z$ and FM spin order and
(d)~AO$z$ and AF spin order.
Image is reproduced from Ref. \cite{Hyper}.
}
\label{fig:good}
\end{figure}

From the derivation of the Kugel-Khomskii model in Sec. \ref{sec:KK},
we have seen that pairs of directional orbitals on neighboring ions
$\{|i\zeta_{\gamma}\rangle,|j\zeta_{\gamma}\rangle\}$ favor AF spin
order while pairs of orthogonal orbitals such as
$\{|i\zeta_{\gamma}\rangle,|j\xi_{\gamma}\rangle\}$ favor FM spin order.
This is generalized to classical Goodenough-Kanamori rules (GKR)
\cite{Goode} that state that AF spin order is accompanied by FO order,
while FM spin order is accompanied by AO order, see
Figs. \ref{fig:good}(a) and \ref{fig:good}(b). Indeed, these rules
emphasizing the complementarity of spin-orbital correlations are
frequently employed to explain the observed spin-orbital order in
several systems, particularly in those where spins are large, like in
CMR manganites \cite{Dag01}.
They agree with the general structure of spin-orbital superexchange in
the Kugel-Khomskii model where it is sufficient to consider the
flavor-conserving hopping between pairs of directional orbitals
$\{|i\zeta_{\gamma}\rangle,|j\zeta_{\gamma}\rangle\}$. The excited
states are then double occupancies in one of the directional orbitals
while no effective interaction arises for two parallel spins
(in triplet states),
so the superexchange is AF. In contrast, for a pair of orthogonal
orbitals, e.g. $\{|i\zeta_{\gamma}\rangle,|j\xi_{\gamma}\rangle\}$, two
different orbitals are singly occupied and the FM term is stronger than
the AF one as the excitation energy is lower. Therefore, configurations
with AO order support FM spin order.

The above complementarity of spin-orbital order is frustrated by
interorbital hopping, or may be modified by spin-orbital entanglement,
see below. In such cases the order in both channels could be the same,
either FM/FO, see Fig. \ref{fig:good}(c), or AF/AO,
see Fig. \ref{fig:good}(d). Again, when different orbitals are
occupied in the excited state, the spin superexchange is weak FM
and when the same orbital is doubly occupied, the spin superexchange
is stronger and AF. The latter AF exchange coupling dominates because
antiferromagnetism, which is due to the Pauli principle, does not
have to compete here with ferromagnetism. On the contrary, FM exchange
is caused by the energy difference $\propto\eta$ between triplet and
singlet excited states with two different orbitals occupied.

The presented modification of the GKR is of importance in alkali
$R$O$_2$ hyperoxides~{($R$=K,Rb,Cs)} \cite{Hyper}. The JT effect is
crucial for this generalization of the GKR --- without it large
interorbital hopping orders the $T^x$-orbital-mixing pseudospin
component instead of the $T^z$ component in a single plane. More
generally, such generalized GKR can arise whenever the orbital order
on a bond is not solely stabilized by the same spin-orbital
superexchange interaction that determines the spin exchange. On a
geometrically frustrated lattice, another route to this behavior can
occur when the ordered orbital component preferred by superexchange
depends on the direction and the relative strengths fulfill certain
criteria.

\subsection{Spin-orbital entanglement}
\label{sec:enta}
\index{spin-orbital entanglement}

A quantum state consisting of two different parts of the Hilbert space
is entangled if it cannot be written as a product state. Similar to it,
two operators are entangled if they give entangled states, i.e., they
cannot be factorized into parts belonging to different subspaces.
This happens precisely in spin-orbital models and is the source of
spin-orbital entanglement \cite{Ole12}.

To verify whether entanglement occurs it suffices to compute and
analyze the spin, orbital and spin-orbital (four-operator)
correlation functions for a bond $\langle ij\rangle$ along $\gamma$
axis, given respectively by
\begin{eqnarray}
\label{ss}
S_{ij}\!& \equiv & \frac{1}{d}\;\sum_n
\lla n\big|{\vec S}_i \cdot {\vec S}_j \big|n\rra\,, \\
\label{tt}
T_{ij}\! & \equiv & \frac{1}{d}\;\sum_n\lla n\big|
({\vec T}_i \cdot {\vec T}_j)^{(\gamma)}\big|n\rra\,,\\
\label{st}
C_{ij}\!& \equiv & \frac1d \sum_n \lla n|
(\vec S_i\cdot\vec S_j-S_{ij})(\vec T_i\cdot\vec T_j-T_{ij})^{(\gamma)} |n\rra
 \\
\!& = & \!\frac{1}{d}\;\sum_n
\lla n\big| ({\vec S}_i \cdot {\vec S}_j)
     ({\vec T}_i \cdot {\vec T}_j)^{(\gamma)} \big|n\rra
-\!\frac{1}{d}
\sum_n\lla n\big|{\vec S}_i \! \cdot \! {\vec S}_j \big|n\rra
\frac{1}{d}
\sum_m\lla m\big|({\vec T}_i \! \cdot \! {\vec T}_j)^{(\gamma)}
|m\rra\,, \nonumber
\end{eqnarray}
where $d$ is the ground state degeneracy, and the pseudospin scalar
product in Eqs. (\ref{tt}) and (\ref{st}) is relevant for a model with
active $t_{2g}$ orbital degrees of freedom. As a representative example
we evaluate here such correlations for a 2D spin-orbital model derived
for NaTiO$_2$ plane \cite{Nor08}, with the local constraint (\ref{cond1})
as in LaTiO$_3$; other situations with spin-orbital entanglement are
discussed in Ref. \cite{Ole12}.

\begin{figure}[t!]
\includegraphics[width=7.4cm]{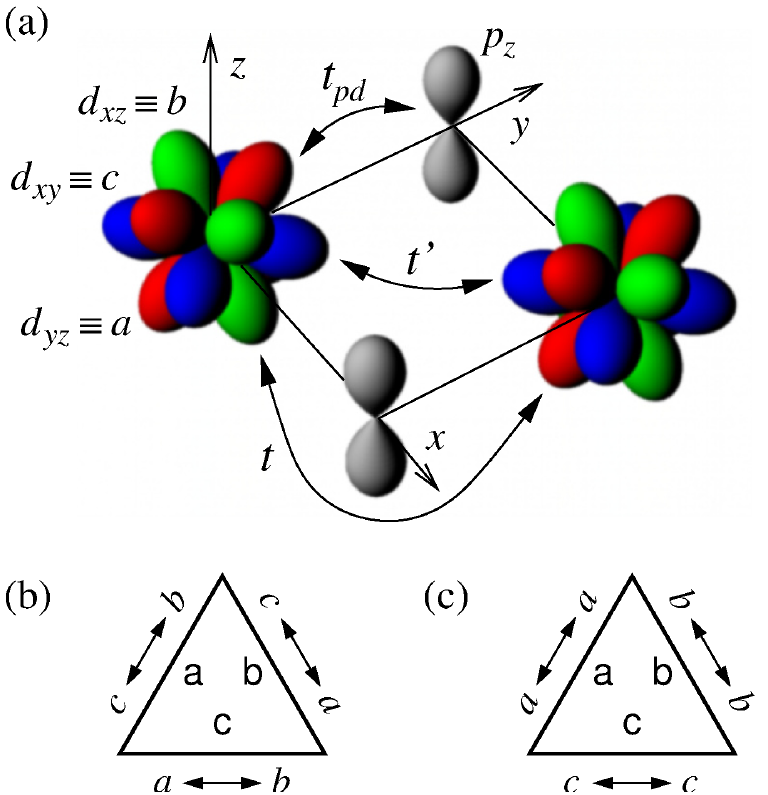}
\hskip .7cm
\includegraphics[width=8.0cm]{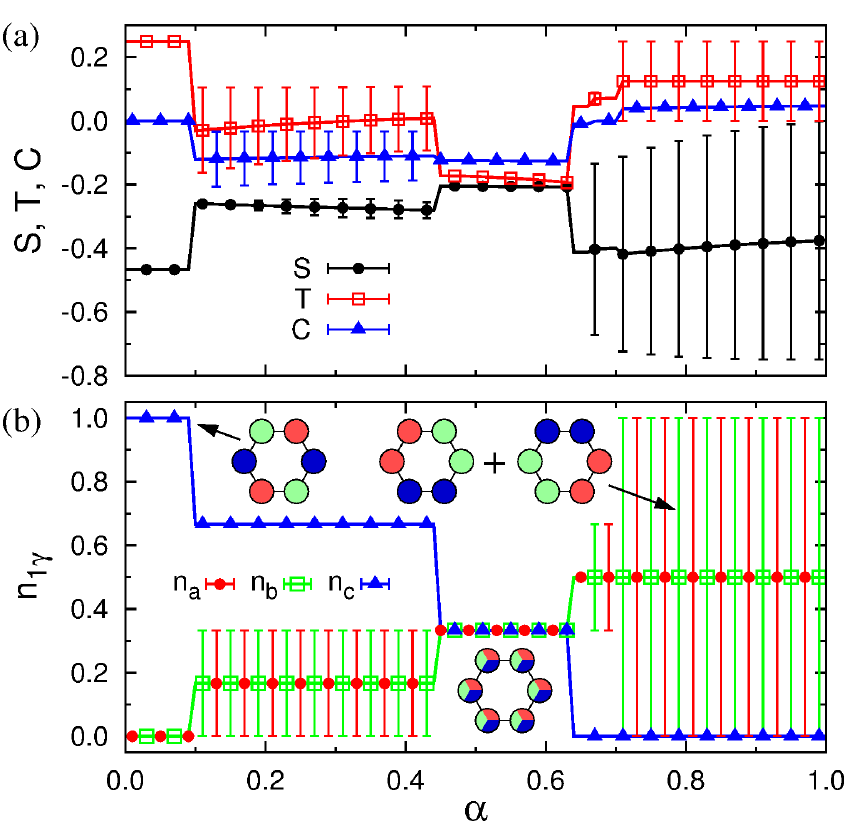}
\caption{
Left ---
(a) Hopping processes between $t_{2g}$ orbitals along a bond parallel
to the $c$ axis in NaTiO$_2$:
(i) $t_{pd}$ between Ti($t_{2g}$) and O($2p_z$) orbitals ---
two $t_{pd}$ transitions define an effective hopping $t$, and
(ii) direct $d-d$ hopping $t'$.
The $t_{2g}$ orbitals shown by different color are labeled as $a$, $b$,
and $c$, see Eq. (\ref{t2g}).
The bottom part gives the hopping processes along $\gamma=a,b,c$ axes
in the triangular lattice that contribute to Eq. (\ref{jiri}):
(b) superexchange and
(c) direct exchange.
Right ---
Ground state for a free hexagon as a function of $\alpha$ (\ref{jiri}):
(a) bond correlations --- spin $S_{ij}$ Eq. (\ref{ss}) (circles),
orbital $T_{ij}$ Eq. (\ref{tt}) (squares), and
spin--orbital $C_{ij}$ Eq. (\ref{st}) (triangles);
(b)~orbital electron densities $n_{1\gamma}$ at a representative site
$i=1$ (left-most site): $n_{1a}$ (circles), $n_{1b}$ (squares),
$n_{1c}$ (triangles). The insets indicate the orbital configurations
favored by the superexchange ($\alpha=0$),
by mixed $0.44<\alpha<0.63$, and
by the direct exchange ($\alpha=1$).
The vertical lines indicate an exact range due to the degeneracy.
Images are reproduced from Ref. \cite{Cha11}.
}
\label{fig:bond}
\end{figure}

To explain the physical origin of the spin-orbital model for NaTiO$_2$
\cite{Nor08} we consider a representative bond along the $c$ axis shown
in Fig. \ref{fig:bond}. For the realistic parameters of NaTiO$_2$ the
$t_{2g}$ electrons are almost localized in $d^1$ configurations of
Ti$^{3+}$ ions, hence their interactions with neighboring sites can be
described by the effective superexchange and kinetic exchange processes.
Virtual charge excitations between the neighboring sites,
$d_i^1d_j^1\rightleftharpoons d_i^2d_j^0$, generate magnetic
interactions which arise from two different hopping processes for
active $t_{2g}$ orbitals:
($i$) the effective hopping $t=t_{pd}^2/\Delta$ which occurs via oxygen
$2p_z$ orbitals with the charge transfer excitation energy $\Delta$, in
the present case along the 90$^{\circ}$ bonds, and
($ii$) direct hopping $t'$ which couples the $t_{2g}$ orbitals along
the bond and give kinetic exchange interaction, as in the Hubbard model
(\ref{J}). Note that the latter processes couple orbitals with the same
flavor, while the former ones couple different orbitals (for this
geometry) so the occupied orbitals may be interchanged as a result of a
virtual charge excitation --- these processes are shown in
Fig. \ref{fig:bond}.

The effective spin-orbital model considered here reads \cite{Nor08},
\index{spin-orbital superexchange!for NaTiO$_2$}
\begin{equation}
\label{jiri}
{\cal H} = J \left\{ (1 - \alpha) \; {\cal H}_s
                 + \sqrt{(1 - \alpha) \alpha} \; {\cal H}_m
                 + \alpha \; {\cal H}_d \right\}\,.
\end{equation}
The parameter $\alpha$ in Eq. (\ref{jiri}) is given by the hopping
elements as follows,
\begin{equation}
\label{alpha}
\alpha=\frac{t'^2}{t^2+t'^2},
\end{equation}
and interpolates between the superexchange ${\cal H}_s$ ($\alpha=0$)
and kinetic exchange ${\cal H}_d$ ($\alpha=1$), while in between mixed
exchange contributes as well; these terms are explained in Ref.
\cite{Nor08}. This model is considered here in the absence of Hund's
exchange $\eta$ (\ref{eta}), i.e., at $\eta=0$.
One finds that all the orbitals contribute equally in the entire range
of $\alpha$, and each orbital state is occupied at two out of six sites
in the entire regime of $\alpha$, see Fig. \ref{fig:bond}. The orbital
state changes under increasing $\alpha$ and one finds four distinct
regimes, with abrupt transitions between them. In the superexchange
model ($\alpha=0$) there is precisely one orbital at each site which
contributes, e.g. $n_{1c}=1$ as the $c$ orbital is active along both
bonds. Having a frozen orbital configuration, the orbitals decouple
from spins and the ground is disentangled, with $C_{ij}=0$, and one
finds that the spin correlations $S_{ij}=-0.4671$, as for the AF
Heisenberg ring of $L=6$ sites. Orbital fluctuations increase gradually
with increasing $\alpha$ and this results in finite spin-orbital
entanglement $C_{ij}\simeq -0.12$ for $0.10<\alpha<0.44$;
simultaneously spin correlations weaken to $S_{ij}\simeq -0.27$.

In agreement with intuition, when $\alpha=0.5$ and all interorbital
transitions shown in Fig.~\ref{fig:bond} have equal amplitude, there
is large orbital mixing which is the most prominent feature in the
intermediate regime of $0.44<\alpha<0.63$. Although spins are coupled
by AF exchange, the orbitals fluctuate here strongly and reduce
further spin correlations to $S_{ij}\simeq -0.21$. The orbital
correlations are negative, $T_{ij}<0$, the spin-orbital entanglement
is finite, $C_{ij}\simeq -0.13$, and the ground state is unique ($d=1$).
Here all the orbitals contribute equally and $n_{1\gamma}=1/3$ which
may be seen as a precursor of the spin-orbital liquid state which
dominates the behavior of the triangular lattice. The regime of larger
values of $\alpha>0.63$ is dominated by the kinetic exchange in Eq.
(\ref{jiri}), and the ground state is degenerate with $d=2$
\cite{Cha11}, with strong scattering of possible electron
densities $\{b_{i\gamma}\}$, see Fig.~\ref{fig:bond}. Weak entanglement
is found for $\alpha>0.63$, where $C_{ij}\simeq\neq 0$. Summarizing,
except for the regimes of $\alpha<0.09$ and $\alpha>0.63$ the ground
state of a single hexagon is strongly entangled, i.e., $C_{ij}<-0.10$,
see Fig.~\ref{fig:bond}.

\subsection{Fractionalization of orbital excitations}
\label{sec:omo}

As a rule, even when spin and orbital operators disentangle in the
ground state, some of the
excited states are characterized by spin-orbital entanglement. It
is therefore even more subtle to separate spin-orbital degrees of
freedom to introduce orbitons as independent orbital excitations,
in analogy to the purely orbital model and the result presented in
Fig. \ref{fig:owa} \cite{Woh11}. This problem is not yet completely
understood and we show here that in a 1D spin-orbital model the
orbital excitation fractionalizes into freely propagating spinon and
orbiton, in analogy to charge-spinon separation in the 1D $t$-$J$
model.
\index{spinon-orbiton separation}
\index{$t$-$J$ model}

\begin{figure}[t!]
\bec
\includegraphics[width=12cm,clip=true]{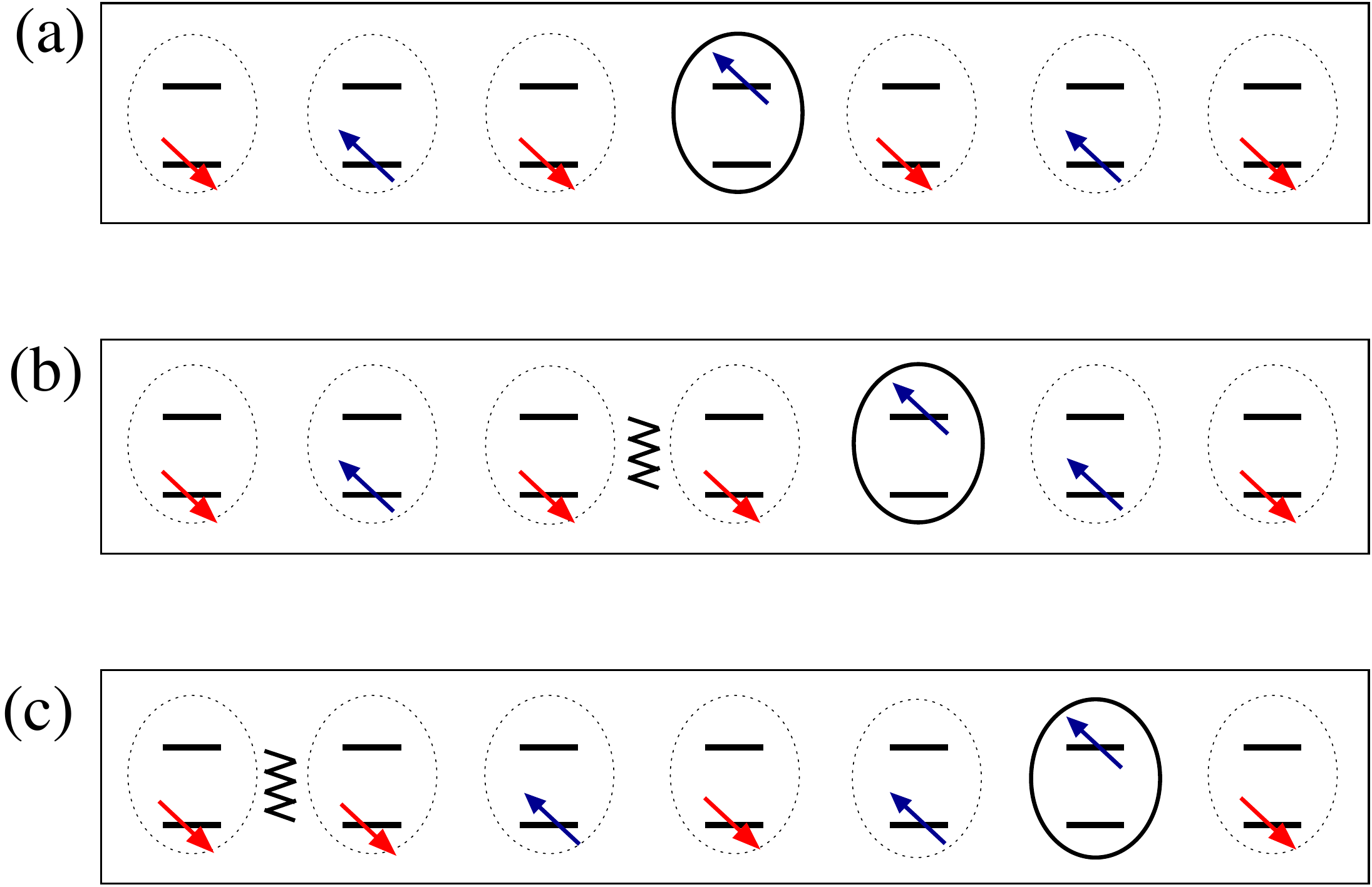}
\enc
\caption{
Schematic representation of the orbital motion and the spin-orbital
separation in a 1D spin-orbital model. The first hop of the excited
state (a)$\to$(b)  creates a spinon (wavy line) that moves via spin
exchange $\propto J$. The next hop (b)$\to$(c) gives an ''orbiton''
freely propagating as a ''holon'' with an effective hopping $t\sim J/2$.
Image is reproduced from Ref. \cite{Woh11}.}
\label{fig:sos}
\end{figure}

While a hole doped to the FM chain propagates freely, it creates a
spinon and a holon in an AF background described by the $t$-$J$ model.
A similar situation occurs
for an orbital excitation in AF/FO spin-orbital chain \cite{Woh11}.
An orbital excitation may propagate through the system only after
creating a spinon in the first step, see Figs. \ref{fig:sos}(a) and
\ref{fig:sos}(b). The spinon itself moves via spin flips $\propto J>t$,
faster than the orbiton, and the two excitations get well separated,
see Fig. \ref{fig:sos}(c). The orbital-wave picture of
Sec.~\ref{sec:orbi},
on the other hand, would require the orbital excitation to move without
creating the spinon in the first step. Note that this would be only
possible for imperfect N\'eel AF spin order. Thus one concludes that
the symmetry between spin and orbital sector is broken also for this
reason and orbitals are so strongly coupled to spin excitations in
realistic spin-orbital models with AF/FO order that the mean field
picture separating these two sectors of the Hilbert space breaks down.

\section{$t$-$J$-like model for ferromagnetic manganites}
\label{sec:tJ}

Even more complex situations arise when charge degrees of freedom are
added to spin-orbital models. The spectral properties of such models
are beyond the scope of this discussion but we shall only point out
that macroscopic doping changes radically spin-orbital superexchange
by adding to it ferromagnetic exchange triggered by $e_g$ orbital
liquid realized in hole doped manganites. As a result, the CMR effect
is observed and the spin order changes to FM \cite{Dag01}.

\begin{figure}[t!]
\bec
\includegraphics[width=15cm,clip=true]{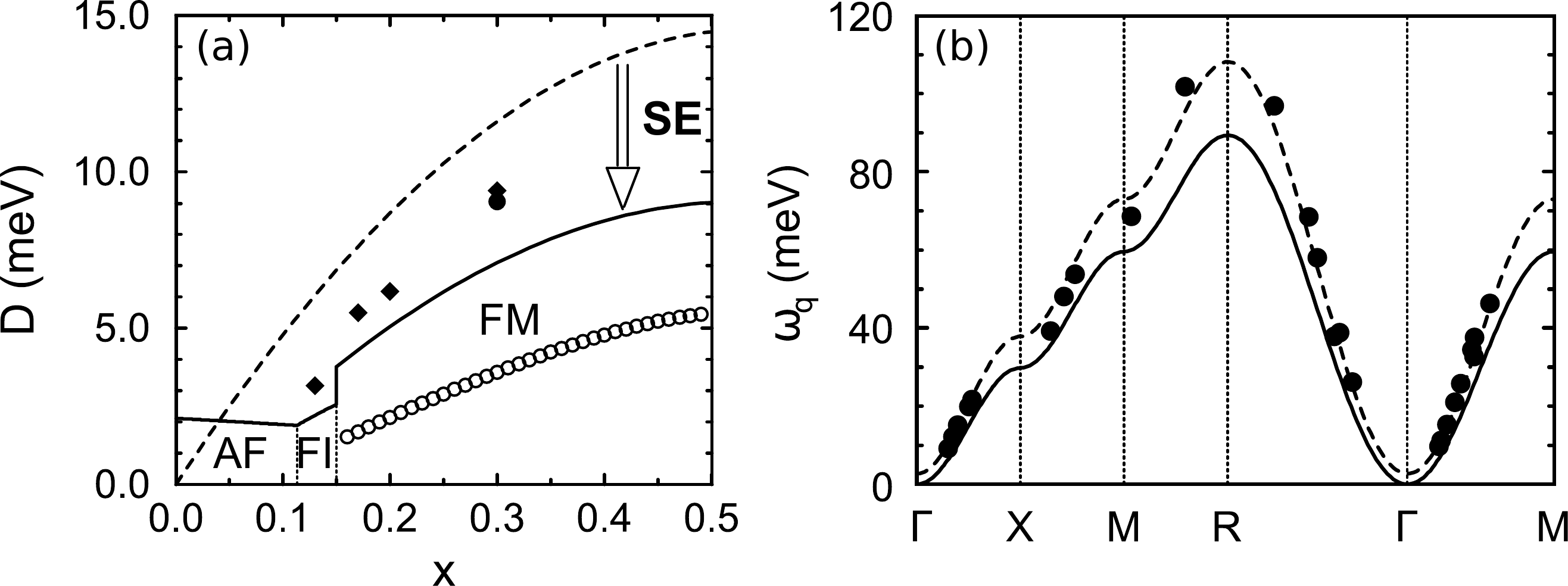}
\enc
\caption{
Theoretical predictions for magnon spectra in the FM metallic phase
in manganites:
\hfill\break
(a) spin-wave stiffness $D$ (solid line) as a function of hole doping
$x$ given by double exchange (dashed) reduced by superexchange (SE) for:
$A$-AF, FM insulating (FI), and FM metallic (FM) phases, and
experimental points for La$_{1-x}$Sr$_x$MnO$_3$ (diamonds) and
La$_{0.7}$Pb$_{0.3}$MnO$_3$ (circle); empty circles for the hypothetical
AO $|\pm\rangle$ state unstable against the $e_g$ orbital liquid;
\hfill\break
(b)~magnon dispersion $\omega_{\vec q}$ obtained at $x=0.30$ (solid line)
and the experimental points for La$_{0.7}$Pb$_{0.3}$MnO$_3$ \cite{Fer98}
(circles and dashed line). \hfill\break
Parameters: $U=5.9$, $J_H^e=0.7$, $t=0.41$, all in eV.
Images are reproduced from Ref. \cite{Ole02}.}
\label{fig:de}
\end{figure}

Similar to the spin case, the orbital Hubbard model Eq. (\ref{egHub})
gives at large $\bar{U}\gg t$ the $e_g$ $t$-$J$ model \cite{Ole02},
i.e., $e_g$ electrons may hop only in the restricted space without
doubly occupied $e_g^2$ sites. The kinetic energy is gradually released
with increasing doping $x$ in doped manganese oxides
La$_{1-x}A_x$MnO$_3$, with $A=$Sr,Ca,Pb, which is a driving
mechanism for effective FM interaction generated by the kinetic energy
$\propto{\tilde H}_t^{\uparrow}(e_g)$ in the double exchange
\cite{Dag01}. It competes with AF exchange which eventually becomes
frustrated in FM metallic phase, arising typically at sufficient hole
doping $x\simeq 0.17$. The evolution of magnetic order with increasing
doping results from the above frustration: at low doping $x\sim 0.1$ AF
spin order becomes stable and first changes to FM insulating phase, see
Fig.~\ref{fig:de}(a). Only at larger doping $x$, an insulator-to-metal
transition takes place which explains the CMR effect \cite{Dag01}.

In the FM metallic phase the magnon excitation energy is derived from
manganite $t$-$J$ model and consists of two terms \cite{Ole02}:
($i$) superexchange being AF for the orbital liquid and
($ii$)~FM~double exchange $J_{\rm DE}$, proportional to the kinetic
energy of $e_g$ electrons (\ref{Hxz}),
\index{$t$-$J$ model}
\index{double exchange}
\beq
\label{Jde}
J_{\rm DE}=\frac{1}{2z{\cal S}^2}\,
\left|\lla{\tilde H}_t^{\uparrow}(e_g)\rra\right|\,.
\eeq
Here $z$ is the number of neighbors ($z=6$ for the cubic lattice), and
$2{\cal S}=4-x$ is the average spin in a doped manganese oxide. The
kinetic energy $\left|\lla{\tilde H}_t^{\uparrow}(e_g)\rra\right|$
measures directly the band narrowing due to the strong correlations in
the $e_g$ orbital liquid. This explains why the spin-wave stiffness $D$
is:
($i$) reduced by the frustrating AF superexchange $J_{\rm SE}$ but
($ii$) increases proportionally to the hole doping $x$ in the FM
metallic phase, see Fig. \ref{fig:de}(a). As a result, the magnon
dispersion in the FM metallic phase is given by,
\beq
\label{fmm}
\omega_{\vec q}=
\left(J_{\rm DE}-J_{\rm SE}\right)z{\cal S}^2(1-\gamma_{\vec q}),
\eeq
where $\gamma_{\vec q}=\frac{1}{z}\sum_{{\vec\delta}}
e^{i{\vec q}\cdot{\vec\delta}}$, and ${\vec\delta}$ is a vector which
connects the nearest neighbors.

An experimental proof that indeed the $e_g$ orbital liquid is
responsible for isotropic spin excitations in the FM metallic phase
of doped manganites we show the magnon spectrum observed in
La$_{0.7}$Pb$_{0.3}$MnO$_3$, with rather large stiffness constant
$D=7.25$ meV, see Fig. \ref{fig:de}(b). Note that $D$ would be much
smaller in the phase with AO order of $|\pm\rangle$ orbitals
(\ref{pm}). Summarizing, the isotropy of the spin excitations in the
simplest manganese oxides with FM metallic phase is naturally explained
by the \textit{orbital liquid} state of disordered $e_g$ orbitals.

\section{Conclusions and outlook}

Spin-orbital physics is a very challenging field in which only certain
and mainly classical aspects have been understood so far. We have
explained the simplest principles of spin-orbital models deciding about
the physical properties of strongly correlated transition metal oxides
with active orbital degrees of freedom. In the correlated insulators
exchange interactions are usually frustrated and this frustration is
released by certain type of spin-orbital order, with the
complementarity of spin and orbital correlations at AF/FO or FM/AO
bonds, as explained by the Goodenough-Kanamori rules~\cite{Goode}.

One of the challenges is spin-orbital entanglement which becomes
visible both in the ground and excited states. The coherent excitations
such as magnons or orbitons are frequently not independent and also
composite spin-orbital excitations are possible. Such excitations are
not yet understood, except for some simplest cases as e.g. the 1D
spin-orbital model with SU(4) symmetry where all these excitations are
on equal footing and contribute to the entropy in the same way
\cite{su4}. Such a perfect symmetry does not occur in nature however,
and the orbital excitations are more complex due to finite Hund's
exchange interaction and, at least in some systems, orbital-lattice
couplings. They may be a decisive factor explaining why
spin-orbital liquids do not occur in certain models. For the same
reason in the absence of geometrical frustration, the orbital liquid
seems easier to obtain than the spin liquid.

\begin{figure}[t!]
\bec
\includegraphics[width=15cm,clip=true]{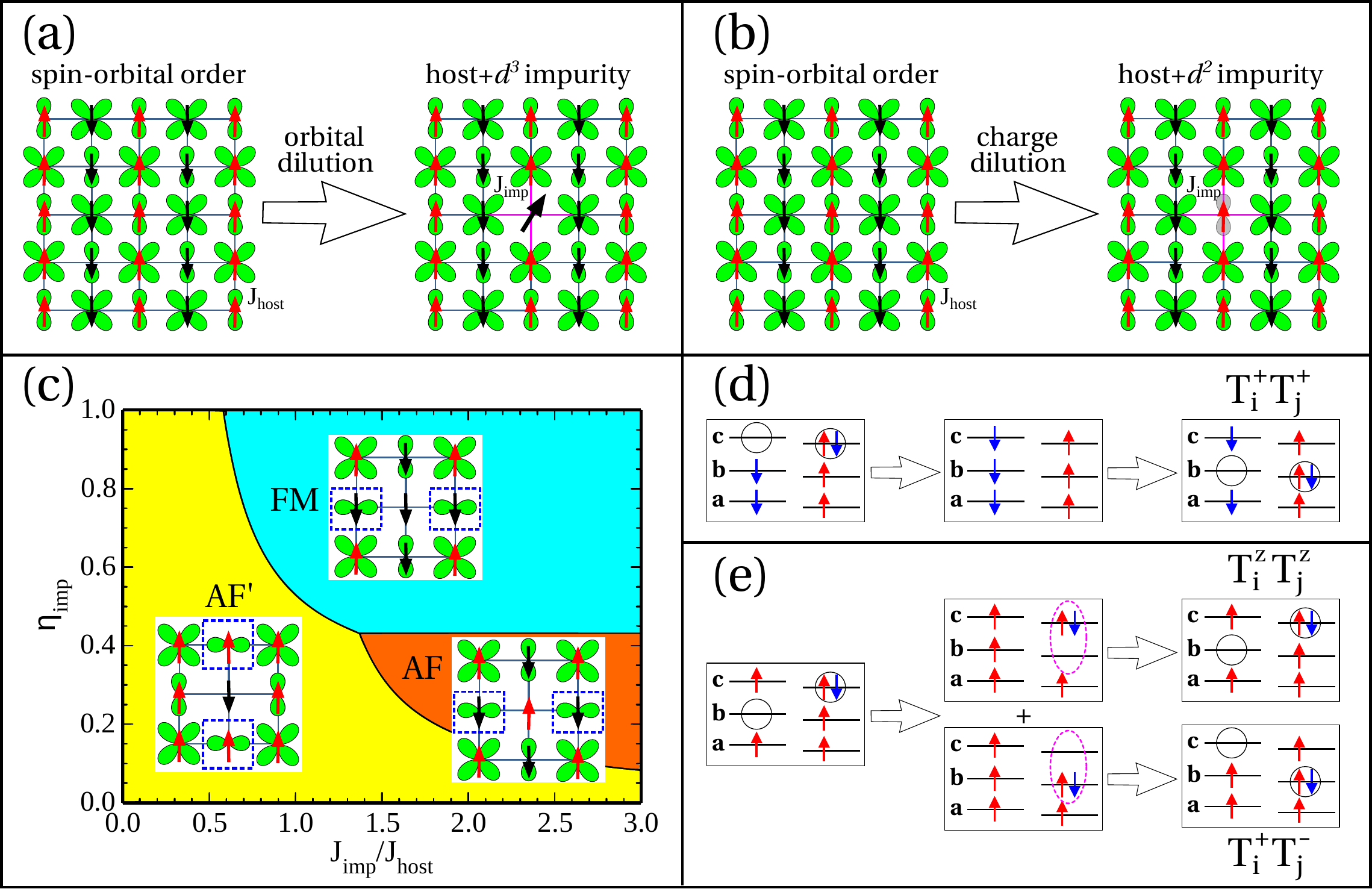}
\enc
\caption{
Top --- Doping by transition metal ions in an $ab$ plane with
$C$-AF/$G$-AO order of $\{a,c\}$ orbitals found in $d^4$ Mott
insulators (ruthenates) with:
(a) orbital dilution by the $d^3$ impurity with $S=3/2$ spin, and
(b) charge dilution by the $d^2$ impurity with $S=1$ spin.
Host $S=1$ spins (red/black arrows) interact by $J_{\rm host}$ and
doublons in $a$ ($c$) orbitals shown by green symbols. Here doping
occurs at $a$ doublon site and spins are coupled by $J_{\rm imp}$
along hybrid (red) bonds. \hfill\break
Bottom ---
(c) phase diagram for a single $d^3$ impurity replacing a doublon in
$c$ orbital in the $C$-AF host \cite{Brz15}, with changes in the
orbital order indicated by dashed boxes (note $a\rightarrow b$ orbital
flips);
(d-e) orbital fluctuations promoted on $d^2$--$d^4$ hybrid bonds with
(d) AF and
(e)~FM spin correlations.
In the latter case (e) the doublons at two orbitals are coupled in
excited states (doublon and hole in ovals), and
one obtains orbital flips $\propto T_i^-T_j^+$
accompanied by Ising terms $\propto T_i^zT_j^z$,
while  double excitations $\propto T_i^+T_j^+$ occur on AF bonds (d)
even in the absence of Hund's exchange and are amplified by finite
$\eta$.
Image is reproduced from Ref. \cite{Brz17}.
}
\label{fig:prx}
\end{figure}
\index{spin-orbital superexchange!at orbital dilution}
\index{spin-orbital superexchange!at charge dilution}

Doping of spin-orbital systems leads to very rich physics with phase
transitions induced by moving charge carriers, as for instance in the
well known example of the CMR manganites. Yet, the holes doped to the
correlated insulators with spin-orbital order may be of quite different
nature. Charge defects may prevent the holes from coherent propagation
\cite{Ave15} and as a result the spin-orbital order will persist to
rather high doping level.

Recently doping by transition metal ions with different valence was
explored \cite{Brz15} --- in such $t_{2g}$ systems local or global
changes of spin-orbital order result from the complex interplay of
spin-orbital degrees of freedom at \textit{orbital dilution}, see
Fig.~\ref{fig:prx}(a). In general, the observed order in the doped
system will then depend on the coupling between the ions with different
valence compared with that within the host $J_{\rm imp}/J_{\rm host}$,
and on Hund's exchange at doped ions $\eta_{\rm imp}$.
Not only a crossover between AF and FM spin correlations is expected
with increasing $\eta_{\rm imp}$, but also the orbital state will
change from inactive orbitals to orbital polarons on the hybrid bonds
with increasing $J_{\rm imp}$, see Fig. \ref{fig:prx}(c). Quite a
different case is given when double occupancies are
replaced by empty orbitals in \textit{charge doping} as shown in
Fig.~\ref{fig:prx}(b). Here orbital fluctuations are remarkably
enhanced by the novel double excitation $\propto T_i^+T_j^+$ terms,
see Figs.~\ref{fig:prx}(d-e). On the one hand, large spin-orbital
entanglement is expected in such cases when Hund's exchange is weak,
while on the other hand the superexchange will reduce to the orbital
model in the FM regime. By mapping of this latter model to fermions
one may expect interesting topological states in low dimension that
are under investigation at present.

\section*{Acknowledgments}
We kindly acknowledge support by Narodowe Centrum Nauki
(NCN, National Science Centre, Poland),
under Project MAESTRO No.~2012/04/A/ST3/00331.



\clearpage

\clearchapter


\end{document}